\documentclass[article,twocolumn,notitlepage,superscriptaddress]{revtex4-1}

\usepackage{setspace}
\usepackage{amsmath}
\usepackage{amssymb}
\usepackage{bbold}
\usepackage{graphicx}
\usepackage{color}
\usepackage[hang,flushmargin]{footmisc} 
\usepackage{amsthm}
\usepackage{comment}

\newcommand{\bx}[0]{\mathbf{x}}
\newcommand{\bv}[0]{\mathbf{v}}
\newcommand{\bp}[0]{\mathbf{p}}
\newcommand{\bw}[0]{\mathbf{w}}
\newcommand{\bd}[0]{\mathbf{d}}
\newcommand{\bW}[0]{\mathbf{W}}

\newcommand{\be}[0]{\mathbf{e}}
\newcommand{\bbf}[0]{\mathbf{f}}

\newcommand{\bR}[0]{\mathbf{R}}
\newcommand{\bY}[0]{\mathbf{Y}}
\newcommand{\bC}[0]{\mathbf{C}}
\newcommand{\bG}[0]{\mathbf{G}}
\newcommand{\bq}[0]{\mathbf{q}}
\newcommand{\bJ}[0]{\mathbf{J}}
\newcommand{\bT}[0]{\mathbf{T}}

\newcommand{\bL}[0]{\mathbf{L}}

\newcommand{\brak}[1]{\left\langle#1\right\rangle}

\newcommand{\tr}[0]{\text{Tr}}

\theoremstyle{plain}

\theoremstyle{definition}

\theoremstyle{remark}

\begin{document}
\renewcommand\abstractname{}
\title{Statistical Mechanics of Transport Processes in Active Fluids II:\\Equations of Hydrodynamics for Active Brownian Particles}
\author{Jeffrey M. Epstein}\email{epstein@berkeley.edu}
\affiliation{Department of Physics, University of California, Berkeley, 94720 USA}

\author{Katherine Klymko}\email{kek2134@berkeley.edu}
\affiliation{Department of Chemistry, University of California, Berkeley, 94720, USA}
\affiliation{Computational Research Division, Lawrence Berkeley National Laboratory, 94720 USA}

\author{Kranthi K. Mandadapu}\email{kranthi@berkeley.edu}
\affiliation{Department of Chemical and Biomolecular Engineering, University of California, Berkeley, 94720 USA}
\affiliation{Chemical Sciences Division, Lawrence Berkeley National Laboratory, 94720 USA}

\begin{abstract}
We perform a coarse-graining analysis of the paradigmatic active matter model, Active Brownian Particles, yielding a continuum description in terms of balance laws for mass, linear and angular momentum, and energy. The derivation of the balance of linear momentum reveals that the active force manifests itself directly as a continuum-level body force proportional to an order parameter-like director field, which therefore requires its own evolution equation to complete the continuum description of the system. We derive this equation, demonstrating in the process that bulk currents may be sustained in homogeneous systems only in the presence of inter-particle aligning interactions. Further, we perform a second coarse-graining of the balance of linear momentum and derive the expression for active or swim pressure in the case of mechanical equilibrium.
\end{abstract}

\maketitle

\section{Introduction}
Active matter models describe non-equilibrium systems that consume and dissipate energy throughout their bulks at the microscopic level even in the absence of gradients~\cite{lauga2009hydrodynamics,ramaswamy2010mechanics,marchetti2013hydrodynamics,bechinger2016active}. This feature distinguishes them from the better-understood class of non-equilibrium systems resulting from gradients in temperature, pressure, chemical potential, or other thermodynamical variables imposed via boundary conditions or fluctuations; indeed, the connection between these two sources of gradients is the content of the fluctuation-dissipation theorems. It is widely expected that studying active systems will provide novel insights into biological processes~\cite{needleman2017active}, lead to potential technological applications~\cite{manna2017colloidal}, perhaps through tunable rheologies~\cite{hatwalne2004rheology,henkin2014tunable,banerjee2017odd,guillamat2016probing}, and promote the development of new fundamental tools in statistical mechanics~\cite{trefz2016activity,baskaran2009statistical,cates2015motility,takatori2015towards,lau2009fluctuating,speck2015dynamical,paliwal2017non}.

In this paper, we work with systems of Active Brownian Particles (ABPs)~\cite{romanczuk2012active,redner2013structure,fily2012athermal,solon2015activec}, spherical particles subject to Brownian motion and an ``active force'' applied in a direction specified by an angular coordinate that also diffuses. This system has been realized experimentally with colloids selectively coated by light-sensitive catalysts~\cite{palacci2013living,buttinoni2013dynamical,buttinoni2012active} and with asymmetric walkers on a vibrating substrate~\cite{narayan2007long}, and provides a model, certainly simplified, for swarms of bacteria~\cite{cates2010arrested,cates2012diffusive,cates2013active,liu2017motility}. Extensive numerical investigations of the model have also been performed, in which several novel phenomena have been observed in systems of ABPs, including motility induced phase separation (MIPS)~\cite{,redner2013structure,fily2012athermal,tailleur2008statistical,prymidis2016vapour}, spontaneous rectification~\cite{wan2008rectification,di2010bacterial,angelani2010geometrically,ghosh2013self,mallory2014curvature}, and density enhancement near boundaries~\cite{Sol15,marconi2017self}.

There has been significant debate about the possibility and utility of defining continuum-level variables, particularly pressure~\cite{mallory2014anomalous,solon2015pressure,ginot2015nonequilibrium,Tak14,takatori2015towards,fily2017mechanical,marini2017pressure}, for ABPs and related systems. It has been argued that because the distributions of particles in confined regions depend on details of the boundaries, pressure is not a bulk property \cite{Sol15}. On the other hand, a ``swim pressure'' or active stress has been defined and used to successfully predict the onset of MIPS in \cite{Tak14}. Moreover, it has been argued that the issues raised in \cite{Sol15} may be avoided by introducing the concept of an active body force \cite{Yan15}. 
However, the microscopic derivation of the expression for this body force, swim pressure, or active stress is missing, with existing proposals simply writing a virial expression for the active force~\cite{yang2014aggregation,Tak14,speck2016ideal}. We address the issue of these microscopic derivations systematically by performing a coarse-graining of the microscopic ABP equations of motion in the style of Irving and Kirkwood~\cite{Irv50,steffenoni2017microscopic} to yield a continuum description~\cite{Puglisi2017,Note1}. 

The Irving-Kirkwood procedure \cite{Irv50} was originally developed to reconcile the continuum hydrodynamical balance equations with the atomistic equations of motion. To this end, Irving and Kirkwood found expressions for the stress tensor and heat flux vector in terms of molecular positions and momenta, thereby providing a microscopic basis for the continuum variables appearing in hydrodynamic theories. Such expressions provide a basis for understanding transport properties, such as viscosity and thermal conductivity, from molecular simulations. A systematic extension of the Irving-Kirkwood procedure to active matter systems like ABPs, as in this work, clarifies the dependence of emergent continuum properties on molecular variables in out-of-equilibrium systems, and facilitates the study of emergent rheological and transport properties of active fluids and suspensions.

This work on ABPs may be considered a continuation of earlier work presented in~\cite{Kly17} on rotary dumbbell particles driven by active torques. To this end, the current work focuses on including active convective forces and analyzing the consequences of this feature. The main contributions of this work are as follows. We
\begin{enumerate}
\item derive balance laws for mass, linear and angular momentum, and energy

\item show that the active force appears naturally in the body force rather than the stress tensor, proportional to an order-parameter-like director field

\item derive equations of motion for this director field, arguing from a mechanical point of view that ABPs can support gradient-free bulk currents only in the presence of inter-particle alignment interactions

\item derive a microscopic expression for heat flux through energy balance

\item provide a derivation of the microscopic expression for swim or active pressure. 
\end{enumerate}

The paper is organized as follows. In section I, we provide a brief description of the microscopic equations of motion for the ABP system. In section II, we discuss the coarse-graining analysis of these equations and the features of the resulting continuum-level balance laws. In section III, we derive equations of motion for the director field whose importance is demonstrated in the balance law for linear momentum, and comment on the implications for bulk currents. In section IV, we perform a second coarse-graining of the linear momentum balance, deriving the active swim pressure. All balance and microscopic expressions for associated fluxes and body contributions are summarized in Table \ref{BalanceTable}. Most derivations are detailed in the Appendix.

\section{Active Brownian Model}
A single ABP in two dimensions is characterized by position $\mathbf{x}_i$, linear momentum $\mathbf{p}_i$, an angular coordinate $\theta_i$, and an internal angular momentum $L_i$ with moment of inertia $I$. The momenta are subject to a drag force/torque and Gaussian noise with strengths $\alpha_p$ and $\alpha_r$ consistent with the fluctuation-dissipation theorems as in standard equilibrium models (although our analysis does not depend on this choice). In addition, an active force $f$ is applied in the direction specified by $\theta_i$ (equivalently, the director or the unit vector $\bd_i(\theta) = \cos\theta_i \be_x + \sin \theta_i \be_y $) and an active torque $\mathbf{\tau}$ is also applied, where $\be_x$, $\be_y$ and $\be_z$ are the unit vectors in $x$, $y$ and $z$ direction, respectively. The particles interact via interparticle forces $\mathbf{f}_{ij}$ and torques $\mathbf{\tau}_{ij}$ assumed to result from a pair potential $u\left(\mathbf{x_i}-\mathbf{x}_j,\theta_i-\theta_j\right)$. The stochastic evolution equations describing the dynamics of ABPs are

\begin{equation}
\begin{aligned}
\frac{d\mathbf{x}_i}{dt}&=\frac{\mathbf{p}_i}{m}\\
\frac{d\mathbf{p}_i}{dt}&=-\frac{\xi_p}{m}\mathbf{p}_i+f\mathbf{d}_i+\sum_j\mathbf{f}_{ij}+\alpha_p\mathbf{W}_i(t)\\
\frac{d\theta_i}{dt}&=\frac{\bL_i}{I}\cdot \mathbf{e}_z\\
\frac{d\bL_i}{dt}&=-\frac{\xi_r}{I}\bL_i+\left(\mathbf{\tau}+\sum_j\mathbf{\tau}_{ij}+\alpha_rZ_i(t)\right)\be_z,
\end{aligned}
\label{eq:sde}
\end{equation}
where the noise terms $\mathbf{W}_i(t)$ and $Z_i(t)$ are zero-mean random processes with second moments given by
\begin{equation}
\begin{aligned}
\brak{W^x_i(t)W^x_j(t')}&=\brak{Z_i(t)Z_j(t')}=\delta_{ij}\xi(t-t'),
\end{aligned}
\end{equation}
{\color{black}where $W^x_i$ are single directional components of $\boldsymbol{W}_i$ and $\xi$ is an even, non-negative function integrating to one. We take the point of view that the random noise is drawn in fact from an ensemble of continuous functions, avoiding the necessity of specifying a stochastic integration scheme, as would be required for true white noise. Taking a limit as the correlation time tends to zero to obtain true stochastic differential equations requires use of the Stratonovich interpretation \cite{Kam80}. We use this convention to demonstrate the decay of the director field in Section \ref{directorsection}.} The linear and angular noise terms are uncorrelated, as are the spatial components of the linear noise. We will use $L_i$ to refer to the magnitude of $\bL_i$, which always points in the $z$-direction, i.e. perpendicular to the plane. Note that we work with an underdamped version of Active Brownian Particles, in contrast to most of the literature, which focuses on the overdamped (non-inertial) limit. The underdamped equations of motion provide us a way to formally treat the balances of linear and angular momenta, and energy.

\section{Balance Laws}
\subsection{Coarse-Graining Approach}
To make contact with a continuum theory, we must derive a set of balance laws from the microscopic equations of motion. To this end, we define fields corresponding to local densities of the basic conserved quantities (mass, linear momentum, angular momentum, and energy) in terms of microscopic variables. Evolution equations for these fields will yield the balance laws. This is essentially the strategy employed by Irving and Kirkwood to derive hydrodynamical equations from microscopic dynamics with ensemble averaging \cite{Irv50}. The difference in our approach is that we define fields without averaging over a statistical ensemble, as also considered by Hardy~\cite{hardy1982formulas}. Thus our balance laws are valid not only on average, but also for any individual trajectory of the system. The ensemble-averaging approach  yields identical results up to the replacement of stochastic noise terms by linear and angular diffusion.

We begin the coarse-graining process by defining the densities as follows:
\begin{equation}
\begin{aligned}
\rho(\bx,t)&=\sum_i m\Delta_i(\bx)\\
\rho\bv(\bx,t)&=\sum_i\bp_i\Delta_i(\bx)\\
\rho \bJ(\bx,t)&=\sum_i \left(\bL_i+\bx_i\times \bp_i\right)\Delta_i(\bx)\\
\rho e(\bx,t)&=\sum_i\left(\frac{\bp_i^2}{2m}+\frac{\bL_i^2}{2I}+\sum_j\frac{u_{ij}}{2}\right)\Delta_i(\bx),
\end{aligned}
\label{fields1}
\end{equation}
where $\Delta_i(\bx)=\Delta(\bx-\bx_i)$, with  $\Delta$ a normalized coarse-graining function assumed to have finite support and to be rotationally invariant.  The results of our work are insensitive to the specific choice of coarse-graining function $\Delta$, since the balance equations are formally the same for any function satisfying the aforementioned properties. In future numerical work, however, the choice of $\Delta$ will be crucial. In particular, $\Delta$ should be chosen to have support large enough to include several particles, so that the continuum fields are relatively smooth, and small enough that these fields are able to capture variation on length scales of interest for the problem at hand. Ideally, the particular setup of the simulation will result in a separation of length scales, with the average interparticle spacing much smaller than the scale on which gradients become relevant; the support of $\Delta$ should be chosen to fall between these two scales.

The interpretation of the coarse-grained fields in the continuum theory is that $\rho$ is the mass density, $\bv$ is the velocity of a continuum point, $\rho \bJ$ is the angular momentum density, and $\rho e$ is the energy density. $\bJ$ always points in the $z$-direction, and its magnitude will be denoted $J$. The same convention will be used throughout the text, with unbolded symbols referring to the magnitude or, equivalently, $z$-component of vectors related to angular momentum.

It will also turn out to be convenient to define the coarse-grained director and noise fields
\begin{equation}
\begin{aligned}
\rho\bd(\bx,t)&=\sum_i\bd_i\Delta_i(\bx)\\
\rho\boldsymbol{\mathcal{W}}(\bx,t)&=\sum_i\alpha_p\bW_i\Delta_i(\bx)\\
\rho\mathcal{Z}(\bx,t)&=\sum_i\alpha_rZ_i\Delta_i(\bx).
\end{aligned}
\label{fields2}
\end{equation}

For fixed $\bx$, each of the fields in Eqs. (\ref{fields1}, \ref{fields2}) is a phase-space function and we may examine its time dependence. Letting $\boldsymbol{\Gamma}$ denote the entire set of phase variables, the time-derivative acts as follows on a phase space function $B$:
\begin{equation}
\frac{d}{dt}B(\boldsymbol{\Gamma}(t))=\left(\frac{d\boldsymbol{\Gamma}}{dt}\cdot\frac{\partial B}{\partial\boldsymbol{\Gamma}}\right)(\boldsymbol{\boldsymbol{\Gamma}}(t)):=\mathcal{F}B(\boldsymbol{\Gamma}(t)).
\end{equation}
For the ABPs, we may read the operator $\mathcal{F}$ directly from the microscopic equations of motion:
\begin{equation}
\begin{aligned}
\mathcal{F}&=\sum_i\frac{\bp_i}{m}\cdot\frac{\partial}{\partial \bx_i}+\frac{L_i}{I}\frac{\partial}{\partial\theta_i}\\
&\hspace{10pt}+\left(-\frac{\xi_p}{m}\bp_i+f\bd_i+\sum_j\bbf_{ij}+\alpha_p\bW_i\right)\cdot\frac{\partial}{\partial \bp_i}\\
&\hspace{10pt}+\left(-\frac{\xi_r}{I}L_i+\tau+\sum_j\tau_{ij}+\alpha_rZ_i\right)\frac{\partial}{\partial L_i}.
\end{aligned}
\end{equation}

With these definitions, we are equipped to determine the forms of the balance laws of the continuum theory. For ease of presentation, we defer the derivation of the balance laws to the appendix and present the laws and microscopic definitions of body contributions and fluxes in Table \ref{BalanceTable}. In the remainder of this section we will discuss the notable features of these balance laws. These are the appearance of the active force in the body force, proportional to the director field $\bd$, the existence of three distinct contributions to the angular momentum, and a modified energy balance compared to that presented in \cite{Kly17}. The mass balance is the standard continuity equation, as derived in Appendix A1. {\color{black} Throughout the table and the rest of this paper we make use of the convective or material derivative $D/Dt=\partial /\partial t+ \mathbf{v}\cdot\partial/\partial \mathbf{x}$.}

\subsection{Linear Momentum Balance: Importance of Director Field}
The first interesting feature appears in the balance of linear momentum; see Appendix A2 for derivation and Table \ref{BalanceTable} for microscopic expressions of body force and stress tensor resulting from the Irving-Kirkwood procedure. In our formulation, the active force $f$ contributes to the body force, rather than to the stress as proposed e.g. in \cite{yang2014aggregation,Tak14,speck2016ideal}. This contribution is mediated by the director field $\bd$, suggesting that this is the relevant order parameter to study alignment order in ABPs, rather than the nematic tensor order parameter, which is invariant under rotation by $\pi$~\cite{prost1995physics,virga1995variational}. We examine the dynamics of the director field $\bd$ in the next section.

We choose to associate to the body force those terms that may not be expressed as divergences. Such a division allows us to interpret the divergence terms as surface-mediated tractions in the momentum balance. Moreover, such a division results in microscopic expressions that reflect the physical nature of the momentum transfer in the system. Note that in principle, the Irving-Kirkwood procedure allows divergence-free tensor field to be added to the stress tensor without altering the balance law. 
We expect this feature to be particularly relevant in future investigations of the behavior of systems of ABPs in various confining geometries.

The notion of swim force or active force as a body force mediated by a director field has been previously discussed in \cite{Yan15}, where the authors demonstrate that non-interacting ABPs undergo sedimentation in the presence of an aligning field but without gravity, and that net active forces resulting from an aligning field may cancel the effects of a body force such as gravity. In fact, the authors point out that inclusion of the active force as a body force resolves the objections of \cite{Sol15} to viewing the pressure as a state function in active systems. Our contribution in this area is to provide a derivation of this role of the active force directly from the microscopic equations of motion.

\subsection{Angular Momentum Balance}
In traditional continuum theories, angular momentum appears only as moment of linear momentum and its balance contains only torques arising from moments of surface tractions and body forces~\cite{landau1986theory}. In these theories, combined with the balance of linear momentum, the balance of angular momentum simply imposes the symmetry of the stress tensor. However, in our system, due both to the internal angular momenta $L_i$ of the individual particles and the effect of coarse-graining, the balance of angular momentum does not follow the traditional form, and contains additional terms requiring distinct microscopic derivation. In order to see this explicitly, we define the following densities:
\begin{equation}
\begin{aligned}
\frac{I}{m}\rho \bw(\bx,t)&=\sum_i\bL_i\Delta_i(\bx)\\
\rho\boldsymbol{\delta}(\bx,t)&=\sum_im\boldsymbol{\delta}_i\Delta_i(\bx)\\
\rho\boldsymbol{\iota}(\bx,t)&=\sum_im\left[\boldsymbol{\delta}_i^2\mathbb{1}-\boldsymbol{\delta}_i\otimes\boldsymbol{\delta}_i\right]\Delta_i(\bx)\\
\rho\boldsymbol{\iota}(\bx,t)\cdot\boldsymbol{\Omega}(\bx,t)&=\sum_i \boldsymbol{\delta}_i\times (\bp_i-m\bv)\Delta_i(\bx),
\end{aligned}
\end{equation}
where $\boldsymbol{\delta}_i=\bx_i-\bx$ measures the positions of particles relative to the material point $\bx$, and $\boldsymbol{\iota}$ and $\boldsymbol{\Omega}$ are a moment of inertia and angular velocity which account for angular momentum about the axis of the continuum point. For large enough coarse-graining volume, we expect $\boldsymbol{\delta}$ to converge rapidly to zero, reflecting that the particles are distributed isotropically with respect to the center of mass. 

The total angular momentum may now be decomposed as follows:
\begin{equation}
\begin{aligned}
\rho \bJ&=\rho \bx\times  \bv+\rho\boldsymbol{\delta}\times  \bv+\rho\boldsymbol{\iota}\cdot\boldsymbol{\Omega}+\frac{I}{m}\rho \bw.
\end{aligned}
\label{totalJ}
\end{equation}
Here, $\rho\bx\times\bv$ is simply the moment of linear momentum. The second and third terms in Eq.~\eqref{totalJ} account for the spatial microstructure~\cite{Note2} of a continuum point, with $\rho\boldsymbol{\delta}\times\bv$ accounting for the difference between the continuum point $\bx$ and the center of mass of the particles in the coarse-graining region centered at $\bx$, and $\rho\boldsymbol{\iota}\cdot\boldsymbol{\Omega}$ for the angular momentum about $\bx$ remaining after transforming to the frame $\bv$ defined by the continuum velocity. The final term $Im^{-1}\rho \bw$ accounts for the internal angular momentum of the ABPs, a second kind of microstructural feature.

The angular momentum derived from the internal spins $\bL_i$ has a moment of momentum that is trivially proportional to the mass density $\rho$, as the moment of inertia of each particle about its center of mass is fixed. The other contributions to angular momentum have moments of inertia which vary according to distinct evolution equations. These are again derived via the Irving-Kirkwood procedure, and are given by
\begin{equation}
\begin{aligned}
\rho\frac{D\boldsymbol{\boldsymbol{\delta}}}{Dt}&=-\frac{\partial}{\partial \bx}\cdot \bR\\
\rho\frac{D\boldsymbol{\iota}}{Dt}&=-\frac{\partial}{\partial \bx}\cdot \bY,
\end{aligned}
\end{equation}
where we have introduced two new fluxes:
\begin{equation}
\begin{aligned}
\bR&=\sum_i\left(\frac{\bp_i}{m}-\bv\right)\otimes m(\bx_i-\bx)\Delta_i(\bx)\\
\bY&=\sum_i\left(\bp_i-m\bv\right)\otimes\left[\boldsymbol{\boldsymbol{\delta}}_i^2\mathbb{1}-\boldsymbol{\boldsymbol{\delta}}_i\otimes\boldsymbol{\boldsymbol{\delta}}_i\right]\Delta_i(\bx).
\end{aligned}
\end{equation}
The latter appeared also in the system of actively rotated dumbbells studied in \cite{Kly17}. See Appendix A3 for derivation.

The decomposition of the total angular momentum also introduces two novel continuum angular velocities $\bw$ and $\boldsymbol{\Omega}$. As opposed to the vorticity of the velocity field $\boldsymbol{\frac{\partial}{\partial\mathbf{x}}}\times\bv$, these are associated to continuum points rather than reflecting properties of the spatial derivatives of linear velocity, and thus are manifestations of the internal microstructure of the continuum points. As such, they require evolution equations of their own, which we derive from the Irving-Kirkwood procedure as
\begin{equation}
\begin{aligned}
\frac{I}{m}\rho \frac{D\bw}{Dt}&=\rho \bG_1+\frac{\partial}{\partial \bx}\cdot \bC_1\\
\rho\boldsymbol{\iota}\cdot\frac{D\boldsymbol{\Omega}}{Dt}&=\rho \bG_2+\frac{\partial}{\partial \bx}\cdot \bC_2+\left(\frac{\partial}{\partial \bx}\cdot \bY\right)\cdot\boldsymbol{\Omega},
\end{aligned}
\end{equation}
where $\bG_1$, $\bG_2$, $\bC_1$, and $\bC_2$ are defined in Table \ref{BalanceTable}. $\bG_1$ and $\bG_2$ are body torques, and $\bC_1$ and $\bC_2$ couple stress tensors. This system then has a total angular momentum balance which is a truly separate balance from linear momentum, and provides an instance of the extension to structured continua of the balance laws proposed by Dahler and Scriven \cite{Dah63}. Note that the components of angular momentum obey balance equations with no explicit interconversion between the components, unless there is a constitutive dependence of the couple stress tensors $\bC_1$ and $\bC_2$ on $\bw$ and $\boldsymbol{\Omega}$ or their gradients. This is in contrast to the previously examined active dumbbell system, where the coupling between the linear and angular momentum balances arises explicitly due to the broken symmetry of the stress tensor \cite{Kly17}.

\subsection{Energy Balance}
Continuing with the Irving-Kirkwood procedure, we obtain the balance of energy, serving as the first law of thermodynamics for active continuous media. See Appendix A4 for derivations and Table \ref{BalanceTable} for the balance law and resulting microscopic expressions for body heating and heat flux. Our choice of decomposition of the energy balance is modified compared to that in \cite{Kly17}. Here, we do not subtract off the term $\boldsymbol{\boldsymbol{\delta}}_i\times\boldsymbol{\Omega}$ from the total momentum $\bp_i$, as done in that earlier work. Our reasoning is that the heating rate $\rho r$ and heat flux $\bq$ are physical quantities, and therefore should be defined in inertial frames, rather than rotating ones. The usefulness of this form remains to be explored.\\
\\
We note that our derivation of the balance laws differs from that of Irving and Kirkwood in that we do not perform ensemble averaging. Performing this additional step does not change the form of our balance laws, and our equations are valid for individual noise trajectories.

\begin{table*}[t]
\boxed{\begin{tabular}{ccc}
Mass & &\(\displaystyle\frac{D\rho}{Dt}=-\rho\left(\frac{\partial}{\partial \bx}\cdot \bv\right)\)\\
\\
Linear Momentum && \(\displaystyle\rho\frac{D\bv}{Dt}=\rho \mathbf{b}+\frac{\partial}{\partial \bx}\cdot \bT\)\\
\\
Angular Momentum && \(\displaystyle \rho\frac{D\bJ}{Dt}=-\rho\left(\frac{\partial}{\partial \bx}\cdot \bR\right)\times \bv+(\bx+\boldsymbol{\boldsymbol{\delta}})\times\left(\rho \mathbf{b}+\frac{\partial}{\partial \bx}\cdot \bT\right)+\rho \bG_1+\frac{\partial}{\partial \bx}\cdot \bC_1+\rho \bG_2+\frac{\partial}{\partial \bx}\cdot \bC_2\)\\
\\
Energy && \(\displaystyle \rho\frac{De}{Dt}=\rho \mathbf{b}\cdot \bv+\frac{\partial}{\partial \bx}\cdot(\bT\cdot \bv)+\rho \bG_1\cdot \bw+\frac{\partial}{\partial \bx}\cdot\left(\bC_1\cdot \bw\right)+\rho r-\frac{\partial}{\partial \bx}\cdot \bq\)\\
\\\hline\\
Body Force && \(\displaystyle \mathbf{b}=-\frac{\xi_p}{m}\bv+f\bd+\boldsymbol{\mathcal{W}}\)\\
\\
Stress Tensor && \(\displaystyle \bT=-\sum_i\frac{\left(\bp_i-m\bv\right)\otimes \left(\bp_i-m\bv\right)}{m}\Delta_i(\bx)-\frac{1}{2}\sum_{ij}b_{ij}\bx_{ij}\otimes \mathbf{f}_{ij}\)\\
\\
Body Torque (Int) && \(\displaystyle \bG_1=-\frac{\xi_r}{m}\bw+\frac{\tau}{m}\be_z+\mathcal{Z}\be_z\)\\
\\
Body Torque (CG) &&\(\displaystyle \rho \bG_2=-\frac{\xi_p}{m}\rho\boldsymbol{\iota}\cdot\boldsymbol{\Omega}-\frac{\xi_p}{m}\boldsymbol{\boldsymbol{\delta}}\times\rho \bv+\sum_i\left(\bx_i-\bx\right)\times\left(\alpha_p\bW_i\right)\Delta_i(\bx)+f\sum_i\left(\bx_i-\bx\right)\times \bd_i\Delta_i(\bx)\)\\
\\
Torque Couple (Int) &&\(\displaystyle \bC_1=-\frac{1}{2}\sum_{ij}b_{ij}\bx_{ij}\tau_{ij}-\sum_i\left(\frac{\bp_i-m\bv}{m}\right)\otimes\left(\bL_i-I\bw\right)\)\\
\\
Torque Couple (CG) &&\(\displaystyle \bC_2=-\sum_i\left(\frac{\bp_i}{m}-\bv\right)\otimes(\bx_i-\bx)\times(\bp_i-m\bv)\Delta_i(\bx)-\frac{1}{2}\sum_{ij}b_{ij}\bx_{ij}\otimes \left(\bx_i-\bx\right)\times \mathbf{f}_{ij}\)\\
\\
Displacement Flux && \(\displaystyle \bR=\sum_i\left(\frac{\bp_i}{m}-\bv\right)\otimes m(\bx_i-\bx)\Delta_i(\bx)\)\\
\\
Body Heating& & \(\displaystyle\rho r=\sum_i\left(-\frac{\xi_p}{m}\bp_i+f\bd_i+\alpha_p\bW_i\right)\cdot\left(\frac{\bp_i}{m}-\bv\right)\Delta_i(\bx)+\sum_i\left(-\frac{\xi_r}{I}\bL_i+\tau+\alpha_rZ_i\right)\cdot \left(\frac{\bL_i}{I}-\bw\right)\Delta_i(\bx)\)\\
\\
Heat Flux & & \(\displaystyle \bq=\sum_i\left(\frac{\bp_i}{m}-\bv\right)\left[\frac{\left(\bp_i-m\bv\right)^2}{2m}+\frac{\left(\bL_i-I\bw\right)^2}{2I}+\sum_j\frac{u_{ij}}{2}\right]\Delta_i(\bx)\)\\
&& \(\displaystyle \hspace{50pt}+\frac{1}{2}\sum_{ij}b_{ij}\left[\left(\bx_{ij}\otimes \mathbf{f}_{ij}\right)\cdot\left(\frac{\bp_i}{m}-\bv\right)+\bx_{ij} \tau_{ij}\left(\frac{\bL_i}{I}-\bw\right)\right]\)
\end{tabular}}
\caption{Balance laws and definitions of necessary fields. ``Int" refers to the contribution to angular momentum from the internal spins $L_i$ while ``CG" refers to the angular momentum associated with the coarse-graining volume. See the discussion of angular momentum balance in the text for definitions of $\boldsymbol{\iota}$, $\boldsymbol{\Omega}$, and $\boldsymbol{\boldsymbol{\delta}}$. The bond function $b_{ij}$ is defined in Appendix A2.}\label{BalanceTable}
\end{table*}

\section{Director Field Dynamics and Bulk Currents}\label{directorsection}
In the previous section, we derived the balance of linear momentum for ABPs, finding that the body force is proportional to a director field $\bd$. Therefore, in order to analyze the motion of the system, it is necessary to derive evolution equations for this field. Moreover, this field represents alignment of particles in the fluid. Unlike the mass, momentum, and energy density fields, the director field does not correspond to a conserved quantity, so its evolution equation does not yield a balance law for the continuum system. However, the director evolution equation can still be derived using the Irving-Kirkwood procedure in an identical fashion as
\begin{equation}
\begin{aligned}
\rho\frac{D\bd}{Dt}&=-\frac{\partial}{\partial \bx}\cdot\sum_i\left(\frac{\bp_i}{m}-\bv\right)\otimes \bd_i\Delta_i(\bx)\\
&\hspace{10pt}+\sum_i\left(\frac{L_i}{I}-w\right)\left(\bd^\perp_i-\bd^\perp\right)\Delta_i(\bx)+\rho w \bd^\perp,
\end{aligned}\label{Dd}
\end{equation}
where $\bw=w\be_z$; see Appendix B1 for detailed derivation. The super-script $\perp$ indicates counterclockwise rotation of the corresponding vector by $\pi/2$.

In a homogeneous system, the first term of the right-hand side of Eq.~\eqref{Dd} vanishes. The third term is orthogonal to $\bd$, hence does not alter the magnitude of the director field. If there is no correlation between the variables $L_i$ and $\bd_i$, the second term approximately vanishes under spatial averaging for large coarse-graining volumes, and angular diffusion will cause any nonzero director field to decay. Indeed, in the absence of aligning interactions between the ABPs, there is no mechanism to generate such correlations, and we see that steady non-zero director fields, and therefore steady flow, is impossible in a homogeneous system. In the presence of aligning interactions, the $L_i$ will in general be correlated with the orientation of the particles relative to the nonzero local director field, when it exists, and the second term will oppose the diffusion, allowing homogeneous steady-state nonzero director fields, and therefore currents.

The connection between currents and inter-particle aligning interactions becomes more evident if we perform an analogous derivation of the director evolution equations using fields defined as noise averages:
\begin{equation}
\begin{aligned}
\bar{\rho}(\bx)&=\brak{\sum_im\Delta_i(\bx)}\\
\bar{\rho} (\bx)\bar{\bv}(\bx)&=\brak{\sum_i \bp_i\Delta_i(\bx)}\\
\bar{\rho} (\bx)\bar{\bd}(\bx)&=\brak{\sum_i\bd_i\Delta_i(\bx)}
\end{aligned}
\end{equation}
and treat the overdamped limit for angular diffusion, where the angular part of the equations of motion is replaced by
\begin{equation}
\frac{d\theta_i}{dt}=\mu\tau+\mu\sum_j\tau_{ij}+\mu\alpha_rZ_i(t),
\end{equation}
with $\mu$ an angular mobility. In this case, the director field evolution is obtained as
\begin{equation}
\begin{aligned}
\bar{\rho}\frac{D\bar{\bd}}{Dt}&=-\frac{1}{2}\mu^2\alpha_r^2\bar{\rho} \bar{\bd}+\mu\tau\bar{\rho} \bar{\bd}^\perp+\boldsymbol{\mathcal{A}}-\frac{\partial}{\partial \bx}\cdot \bJ_d,
\end{aligned}\label{Ddover}
\end{equation}
where
\begin{equation}
\begin{aligned}
\bJ_d&=\brak{\sum_i\left(\bp_i-\bv\right)\otimes \bd_i\Delta_i(\bx)}\\
\boldsymbol{\mathcal{A}}&=\mu\brak{\sum_{ij}\tau_{ij}\bd^\perp_i\Delta_i(\bx)}.
\end{aligned}
\end{equation}
\textcolor{black}{See Appendix B2 for details of the derivation, where we have used the Stratonovich convention.} The first term, proportional to $\bar{\bd}$, represents an exponential decay of the director field due to angular diffusion. The second is a rotation due to the active torque $\tau$. The term $\boldsymbol{\mathcal{A}}$ is a ``body alignment'' vector, and the divergence term accounts for director flux due to particle exchange.

Suppose that the system of ABPs is homogeneous and in a steady state. Then the director evolution equation \eqref{Ddover} reduces to
\begin{equation}
\begin{aligned}
0&=-\mu^2\alpha_r^2\bar{\rho} \bar{\bd}+\mu\tau\bar{\rho} \bar{\bd}^\perp+\boldsymbol{\mathcal{A}}\hspace{1pt}.
\end{aligned}
\end{equation}
In the absence of inter-particle aligning interactions, $\tau_{ij}=0$ and therefore $\boldsymbol{\mathcal{A}}$ is identically zero. Using $\bar{\bd}\cdot\bar{\bd}^\perp=0$, we see that the only solution is $\bar{\bd}=\boldsymbol{0}$. Homogeneity and time-independence also imply that $\mathbf{b}=\boldsymbol{0}$ via the linear momentum balance equation. For large enough coarse-graining volume, $\boldsymbol{\mathcal{W}}$ vanishes, so this in turn implies $\bar{\bv}=\boldsymbol{0}$, demonstrating that nonzero steady-state currents are impossible in homogeneous systems of ABPs without aligning interactions.

In the presence of inter-particle aligning interactions, suppose that $\bar{\bd}\neq \boldsymbol{0}$. The angular noise is symmetric with respect to $\bar{\bd}$, i.e., the distribution of particle directors about $\bar{\bd}$ is symmetric with respect to reflection over $\bar{\bd}$. The particles whose directors point to the left of $\bar{\bd}$ will have a torque with negative component in the $z$ direction, while those pointing to the right a positive torque. Thus, in the sum that defines $\boldsymbol{\mathcal{A}}$, the components of the various $\bd^\perp_i$ in the direction orthogonal to $\bar{\bd}$ will tend to cancel, while the components in the direction of $\bar{\bd}$ will tend to add constructively, giving rise to a nonzero $\boldsymbol{\mathcal{A}}$ in the direction of $\bar{\bd}$, i.e. in opposition to the diffusion-induced decay. Thus, the existence of nonzero steady currents \cite{Toner1995} with $\bar{\bv} = \boldsymbol{0}$ in homogeneous systems of ABPs is possible. {\color{black}We note that this phenomenon has been previously observed in simulations \cite{Vicsek1995,Guillaume2004,Mahault2018}; however, our work presents a continuum interpretation of this phenomenon starting from microscopic equations of motion, and provides predictions for when such currents can arise.}

It is also possible to study systems with boundaries that apply torques to the particles, and in such systems, currents could be expected to appear close to the boundary even in the absence of interparticle aligning interactions. However, our analysis suggests that this effect will persist only in the near-boundary region, and will be negligible deep in the bulk of sufficiently large systems. Noninteracting ABPs in boundary-less regions will not display bulk currents. On the other hand, we expect that in the presence of interparticle alignment, ABPs may display bulk currents that spontaneously break rotational symmetry, even in the absence of boundaries. Addition of boundaries with aligning interactions could then be used to select the direction of bulk currents in arbitrarily large systems.

{\color{black}
\section{Reconciliation with Swim Pressure}
In \cite{Tak14}, Takatori and Brady propose a continuum-level contribution to the stress, called the swim stress. They use the trace of this stress, the swim pressure, to explain simulations showing motility induced phase separation in ABPs. This proposal contrasts with our conclusion that the active force should be accounted for in the body force, rather than the stress, of a continuum theory. The purpose of this section is to resolve this discrepancy. In this section, we will first provide a derivation of the swim pressure in the atomistic setting. Then we will show that this term may be understood as a consequence of the body force via a continuum homogenization procedure in the spirit of \cite{Man12}.

Both the atomistic and continuum analyses presented in this section rely on the notion that pressure should be understood as a force per area that must be applied by an external agent in order to confine a system (which we think of either as a collection of particles or as a continuum body) to a region $\Omega$. We therefore begin by defining representations of the quantities $\boldsymbol{\mathcal{F}}(\mathcal{A})$, the net confinement forces or surface tractions that must be applied to regions $\mathcal{A}\subset\partial\Omega$ of the boundary. Under the assumption that these surface tractions may be understood (on large enough length scales) in terms of a homogeneous pressure $p$, we may express $p$ in terms of confinement forces acting on the near-boundary parts of the system. We then use virial theorems relating bulk and boundary variables in the atomistic and continuum settings to provide a bulk expression for $p$, and recover the expressions for swim pressure as introduced in \cite{Tak14}.

\subsection{Atomistic Representation}
To confine particles to a region $\Omega$, an external agent must apply confining forces $\mathbf{c}_i$. If this force is short-ranged, $\mathbf{c}_i$ is non-zero only for particles close to the boundary. Then we can define $\boldsymbol{\mathcal{F}}(\mathcal{A})$, the net confinement force applied on the region $\mathcal{A}\subset\partial\Omega$ of the boundary:
\begin{equation}
\boldsymbol{\mathcal{F}}(\mathcal{A})=\sum_{i:\mathbf{x}_i\in\mathcal{A}_\Delta}\mathbf{c}_i,\label{fatomistic}
\end{equation}
where $\mathcal{A}_\Delta$ is the neighborhood with radius $\Delta$ around $\mathcal{A}$ for some $\Delta$ larger than the range of the confining force. Clearly this definition should only be used for $\mathcal{A}$ on a scale larger than the typical inter-particle spacing, as particles close to the boundary between any two adjacent regions $\mathcal{A}$ and $\mathcal{A}'$ will be double-counted.

One can define a uniform externally applied pressure $p$ if it is the case that for sufficiently large regions $\mathcal{A}$ of the boundary we have
\begin{equation}
\boldsymbol{\mathcal{F}}(\mathcal{A})\approx-\int_{\mathcal{A}}p\mathbf{n}\,da\label{FAatom},
\end{equation}
with $\mathbf{n}$ the unit normal. If on the other hand the characteristic length scale of $\mathcal{A}$ is much smaller than the smallest scale of boundary curvature (so that $\mathbf{n}$ is roughly constant on $\mathcal{A}$) and letting $\mathbf{x}_0$ be the center of mass of $\mathcal{A}$, we may write
\begin{equation}
\boldsymbol{\mathcal{F}}(\mathcal{A})\cdot\mathbf{x}_0\approx-\int_{\mathcal{A}}p\mathbf{n}\cdot \mathbf{x}\,da.\label{pgeometry}
\end{equation}
If it is also the case that particles in $\mathcal{A}_\Delta$ experience approximately the same constraint force $\mathbf{c}_i$, we have from Eq. (\ref{fatomistic})
\begin{equation}
\boldsymbol{\mathcal{F}}(\mathcal{A})\cdot\mathbf{x}_0\approx\sum_{i:\mathbf{x}_i\in\mathcal{A}_\Delta}\mathbf{c}_i\cdot\mathbf{x}_i.\label{pparticle}
\end{equation}
We are considering in other words the case in which there is an intermediate asymptotic length scale between the particle and continuum scales. If we decompose the entire boundary into regions $\mathcal{A}_k$ of this intermediate scale, we may equate expressions  (\ref{pgeometry}) and (\ref{pparticle}) and sum over $k$ to obtain
\begin{equation}
-\int_{\partial\Omega}p\mathbf{n}\cdot \mathbf{x}\,da\approx\sum_{i}\mathbf{c}_i\cdot\mathbf{x}_i.
\end{equation}
The left-hand side is simply $-pVd$, with $V$ the volume of $\Omega$ and $d$ the dimension of space. We thus obtain
\begin{equation}
p\approx-\frac{1}{Vd}\sum_{i}\mathbf{c}_i\cdot\mathbf{x}_i.\label{atomisticp}
\end{equation}
Note that the summation over $i$ in Eq. (\ref{atomisticp}) runs over all particles, but contributes only for near-boundary particles.

We would like to have an expression for the externally applied pressure in terms of bulk, rather than boundary, variables. In order to obtain such an expression, we appeal to the virial theorem (see Appendix~\ref{app:Virial} for a derivation):
\begin{equation}
\brak{\sum_i\mathbf{f}_i\cdot \mathbf{x}_i+\sum_i\frac{\mathbf{p}_i\cdot \mathbf{p}_i}{m}}=0,
\end{equation}
where brackets indicate ensemble-averaging and the system is assumed to be in a steady state. Here $\mathbf{f}_i$ is the total force, including the constraint forces $\mathbf{c}_i$ applied by the boundary. Separating out the constraint force and using Eq. (\ref{atomisticp}), we have
\begin{equation}
\brak{p}\approx\frac{1}{Vd}\brak{\sum_i(\mathbf{f}_i-\mathbf{c}_i)\cdot \mathbf{x}_i+\sum_i\frac{\mathbf{p}_i\cdot \mathbf{p}_i}{m}}.
\end{equation}

Defining $p_\text{ext}=\brak{p}$ and using the equation (\ref{eq:sde}) for the force on particle $i$, we find
\begin{equation}
\begin{aligned}
p_\text{ext}&\approx-\frac{\xi_pn}{md}\brak{\mathbf{p}_i\cdot \mathbf{x}_i}+\frac{fn}{d}\brak{\mathbf{d}_i\cdot \mathbf{x}_i}+\frac{n}{md}\brak{\mathbf{p}_i\cdot \mathbf{p}_i}\\
&\hspace{20pt}+\frac{1}{Vd}\brak{\frac{1}{2}\sum_{ik}\mathbf{f}_{ik}\cdot \mathbf{x}_{ik}}.\label{pswimatomistic}
\end{aligned}
\end{equation}
The second term is precisely the swim pressure as defined in \cite{Tak14}.

\subsection{Continuum Representation}
In what follows, we show that the atomistic expression (\ref{pswimatomistic}) derived in the previous subsection is consistent with the appearance of the active force in the body force obtained from the Irving-Kirkwood analysis presented earlier. We start by deriving an expression for pressure in a self-contained continuum model in the case of homogeneous surface traction.

To apply a similar argument as used in the atomistic case to the continuum setting, we start with the standard continuum equations of motion, i.e. the mass and momentum balances:
\begin{equation}
\begin{aligned}
\frac{D\rho}{Dt}&=-\rho\frac{\partial}{\partial\mathbf{x}}\cdot \mathbf{v}\\
\rho\frac{D\mathbf{v}}{Dt}&=\rho\mathbf{b}+\frac{\partial}{\partial\mathbf{x}}\cdot\mathbf{T}\label{balances}
\end{aligned}
\end{equation}
with body force $\mathbf{b}$ and stress $\mathbf{T}$. Note that these are the equations of motion that emerged from the Irving-Kirkwood theory. Now we consider a slightly unorthodox formulation of the theory of a continuum body with boundary. Rather than taking the surface traction on an area element $da$ with normal $\mathbf{n}$ to be given by $\mathbf{t}=\mathbf{T}\cdot \mathbf{n}$, we instead view the traction as a part of the body force with support only very close to $\partial\Omega$. Analogous to the confining forces $\mathbf{c}_i$ in the atomistic setting, this component of the body force is considered to be applied by an external agent and is responsible for confining the system. We then decompose the total body force as $\mathbf{b}=\mathbf{b}_c+\mathbf{b}'$, where $\mathbf{b}_c$ is the confining term, vanishing away from $\partial\Omega$, and $\mathbf{b}'$ is the remaining part of the total body force.

Analogous to Eq. (\ref{fatomistic}), we define the quantities $\boldsymbol{\mathcal{F}}(\mathcal{A})$, the net forces applied by the region $\mathcal{A}$ of the boundary, as
\begin{equation}
\boldsymbol{\mathcal{F}}(\mathcal{A})=\int_{\mathcal{A}_\delta}\rho\mathbf{b}_c \, dv,
\end{equation}
where $\mathcal{A}_\delta$ is a neighborhood of radius $\delta$ around $\mathcal{A}$, with $\delta$ large enough to account completely for the non-vanishing constraint body force.

As in Eq. (\ref{FAatom}), we suppose that for any sufficiently large $\mathcal{A}$, we have
\begin{equation}\label{eq:continuum_bodyapprox_0}
\boldsymbol{\mathcal{F}}(\mathcal{A})\approx-\int_{\mathcal{A}}p\mathbf{n}\,da.
\end{equation}
If on the other hand the characteristic length scale of $\mathcal{A}$ is much smaller than the scale of boundary curvature and the scale at which the confinement body force varies, then as in Eqs. (\ref{pgeometry}) and (\ref{pparticle}), we have
\begin{equation}\label{eq:continuum_bodyapprox}
\begin{aligned}
\boldsymbol{\mathcal{F}}(\mathcal{A})\cdot \mathbf{x}_0&\approx\int_{\mathcal{A}_\delta}\rho\mathbf{b}_c\cdot\mathbf{x}\,dv\\
\boldsymbol{\mathcal{F}}(\mathcal{A})\cdot \mathbf{x}_0&\approx-\int_{\mathcal{A}}p\mathbf{n}\cdot \mathbf{x}\,da
\end{aligned}
\end{equation}
where $\mathbf{x}_0$ is the center of mass of $\mathcal{A}$. If we decompose the boundary into disjoint regions $\mathcal{A}_k$ characterized by the intermediate asymptotic length scale that satisfies the assumptions leading to Eqs.~(\ref{eq:continuum_bodyapprox_0}) and (\ref{eq:continuum_bodyapprox}), then we can equate the two expressions in Eqs.~(\ref{eq:continuum_bodyapprox}) and sum over $k$ to obtain
\begin{equation}
\begin{aligned}
-\int_{\partial\Omega}p\mathbf{n}\cdot \mathbf{x}\,da\approx\int_{\partial\Omega_\delta}\rho\mathbf{b}_c\cdot\mathbf{x}\,dv.
\end{aligned}
\end{equation}
Again, the left-hand side is simply $-pVd$, so that we can express the pressure in terms of the confining body force:
\begin{equation}\label{eq:pressure_body}
\begin{aligned}
p&\approx-\frac{1}{dV}\int_{\partial\Omega_\delta}\rho\mathbf{b}_c\cdot\mathbf{x}\,dv.
\end{aligned}
\end{equation}
Because the confining body force vanishes outside of $\partial\Omega_\delta$, we can extend the integral in Eq.~(\ref{eq:pressure_body}) over all space $\mathbb{R}^d$ leading to
\begin{equation}
\begin{aligned}
p&\approx-\frac{1}{dV}\int_{\mathbb{R}^d}\rho\mathbf{b}_c\cdot\mathbf{x}\,dv.\label{pcontinuum}
\end{aligned}
\end{equation}

As in the atomistic setting, we would like an expression of the homogeneous applied pressure $p$ in terms of bulk variables. We therefore appeal now to the continuum version of the virial theorem, which makes use of the balance laws Eq. (\ref{balances}) (see Appendix~\ref{app:Virial} for a derivation):
\begin{equation}
\begin{aligned}
	0&=\int_{\mathbb{R}^d}\rho \mathbf{v}\cdot \mathbf{v}dv+\int_{\mathbb{R}^d}\rho \mathbf{b}\cdot \mathbf{x}\,dv-\int_{\mathbb{R}^d}\tr\left(\mathbf{T}\right)dv,
	\end{aligned}
\end{equation}
where again the system is assumed to be in a steady state. Separating out the constraint part of the body force and using Eq. (\ref{pcontinuum}), we obtain
\begin{equation}\label{eq:pressure_body2}
\begin{aligned}
p&=\frac{1}{Vd}\int_{\mathbb{R}^d}\left[\rho \mathbf{v}\cdot \mathbf{v}+\rho \left(\mathbf{b}-\mathbf{b}_c\right)\cdot \mathbf{x}-\tr\left(\mathbf{T}\right)\right]dv.
\end{aligned}
\end{equation}
Noting that the body force defined in Table \ref{BalanceTable} from the Irving-Kirkwood procedure corresponds to $\mathbf{b}-\mathbf{b}_c$ and the microscopic expression for the stress tensor $\mathbf{T}$, it may be seen that Eq.~(\ref{eq:pressure_body2}) reproduces Eq. (\ref{pswimatomistic}) for the ABP system assuming sufficiently small coarse-graining radius.

Note that in order to make contact with the standard picture in which the traction is given by the normal component of the stress at the boundary, we may assume that the constraint body force is a singular distribution supported on $\partial\Omega$. Then, in the event that the surface traction is homogeneous, we are free to reexpress this distribution as the divergence of an indicator function that is zero outside $\Omega$ and takes a constant value inside $\Omega$. This constant term may then be added to the stress, and appears as an extra pressure. However, this may not be extended to scenarios with inhomogeneities, and therefore any continuum theory of ABPs that is able to deal with alignment and currents will have to incorporate activity via the body force, and not the stress tensor. Furthermore, it will be necessary to solve a coupled problem involving the momentum balance and the non-conservative director evolution equation.}

\section{Conclusion}
In this work, we have performed a coarse-graining analysis of the ABP equations of motion, resulting in a set of balance laws that constitute a continuum description of the system. Although we do not yet have constitutive equations for the the new fields that must be defined, we are able to show the natural appearance of the active force in the body force, to demonstrate the importance of the director field, and to show that aligning interactions are necessary for gradient-free flows or currents. We also derive a balance of energy, which serves as the first law of thermodynamics. This provides a natural setting in which to develop the irreversible thermodynamics of active continuous media driven by convective forces. {\color{black}In addition, molecular dynamics simulations may use these microscopic definitions in order to numerically access emergent viscosity and other transport phenomena.}\\
\\
\noindent\textit{Acknowledgments:}  The authors are indebted to Dibyendu Mandal for useful discussions on the origin and derivation of active pressure. JE was supported by the Department of Defense (DoD) through the National Defense Science \& Engineering Graduate Fellowship (NDSEG) Program. KKM was supported by Director, Office of Science, Office
of Basic Energy Sciences, Chemical Sciences Division, of
the U.S. Department of Energy under contract No. DEAC02-05CH11231.

\newpage
\onecolumngrid
\appendix
\section{Derivation of Balance Laws}
In this appendix, we provide the more detailed derivations of the balance laws presented in Table \ref{BalanceTable}. We begin by defining the field fields corresponding to densities of conserved quantities (mass, linear momentum, angular momentum, and energy):
\begin{equation}
\begin{aligned}
\rho(\bx)&=\sum_i m\Delta_i(\bx)\\
\rho(\bx)\bv(\bx)&=\sum_i\bp_i\Delta_i(\bx)\\
\rho(\bx)\bJ(\bx)&=\sum_i \left(\bL_i+\bx_i\times \bp_i\right)\Delta_i(\bx)\\
\rho(\bx)e(\bx)&=\sum_i\left(\frac{\bp_i^2}{2m}+\frac{\bL_i^2}{2I}+\sum_j\frac{u_{ij}}{2}\right)\Delta_i(\bx).
\end{aligned}
\end{equation}
The angular momentum density may be decomposed into a term due to the moment of linear momentum, another due to the internal angular momenta $L_i$ of the particles, and a third due to the coarse-graining. In order to see this, we define
\begin{equation}
\begin{aligned}
\frac{I}{m}\rho(\bx)\bw(\bx)&=\sum_i\bL_i\Delta_i(\bx)\\
\rho(\bx)\boldsymbol{\boldsymbol{\delta}}(\bx)&=\sum_im(\bx_i-\bx)\Delta_i(\bx)\\
\rho(\bx)\boldsymbol{\iota}(\bx)&=\sum_im\left[(\bx_i-\bx)^2\mathbb{1}-(\bx_i-\bx)\otimes(\bx_i-\bx)\right]\Delta_i(\bx)\\
\rho(\bx)\boldsymbol{\iota}(\bx)\cdot\boldsymbol{\Omega}(\bx)&=\sum_i (\bx_i-\bx)\times (\bp_i-m\bv(\bx))\Delta_i(\bx).
\end{aligned}
\end{equation}
Now using (A2), the angular momentum density may be rewritten as
\begin{equation}
\begin{aligned}
\rho \bJ&=\rho(\bx+\boldsymbol{\boldsymbol{\delta}})\times  \bv+\rho\boldsymbol{\iota}\cdot\boldsymbol{\Omega}+\frac{I}{m}\rho \bw.
\end{aligned}
\end{equation}
The energy may also be decomposed into a term due to the mean velocity, one due to the mean internal angular momentum, and an internal energy term accounting for the potential energy and the deviation of individual particle momenta from the mean. This term is defined as
\begin{equation}
\rho(\bx)\epsilon(\bx)=\sum_i\left(\frac{\left(\bp_i-m\bv(\bx)\right)^2}{2m}+\frac{\left(\bL_i-I\bw(\bx)\right)^2}{2I}+\sum_j\frac{u_{ij}}{2}\right)\Delta_i(\bx).
\end{equation}
Then we have
\begin{equation}
\begin{aligned}
\rho e&=\rho\epsilon+\frac{1}{2}\rho \bv^2+\frac{1}{2}\frac{I}{m}\rho \bw^2.
\end{aligned}
\end{equation}
It will also turn out to be convenient to define the coarse-grained director and noise fields
\begin{equation}
\begin{aligned}
\rho(\bx)\bd(\bx)&=\sum_i\bd_i\Delta_i(\bx)\\
\rho(\bx)\boldsymbol{\mathcal{W}}(\bx)&=\sum_i\alpha_p\bW_i\Delta_i(x)\\
\rho(\bx)\mathcal{Z}(\bx)&=\sum_i\alpha_rZ_i\Delta_i(\bx).
\end{aligned}
\end{equation}
To evaluate the time dependence of these fields, we note that on a phase variable $B$, the time-derivative acts as follows:
\begin{equation}
\frac{d}{dt}B(\Gamma(t))=\mathcal{F}B(\Gamma(t))=\frac{d\Gamma}{dt}\cdot\frac{\partial B}{\partial\Gamma}.
\end{equation}
For the ABPs:
\begin{equation}
\mathcal{F}=\sum_i\frac{\bp_i}{m}\cdot\frac{\partial}{\partial \bx_i}+\left(-\frac{\xi_p}{m}\bp_i+f\bd_i+\sum_j\mathbf{f}_{ij}+\alpha_p\bW_i\right)\cdot\frac{\partial}{\partial \bp_i}+\frac{L_i}{I}\frac{\partial}{\partial\theta_i}+\left(-\frac{\xi_r}{I}L_i+\tau+\sum_j\tau_{ij}+\alpha_rZ_i\right)\frac{\partial}{\partial L_i}.
\end{equation}
Now we are equipped to determine the forms of the balance laws of the continuum theory.

\subsection{Balance of mass}
We have
\begin{equation}
\begin{aligned}
\frac{D\rho}{Dt}&=\frac{\partial\rho}{\partial t}+\left(\bv\cdot\frac{\partial}{\partial \bx}\right)\rho=\mathcal{F}\sum_im\Delta_i(\bx)+\left(\bv\cdot\frac{\partial}{\partial \bx}\right)\rho=\sum_i\bp_i\cdot\frac{\partial}{\partial \bx_i}\Delta_i(\bx)+\left(\bv\cdot\frac{\partial}{\partial \bx}\right)\rho\\
&=-\frac{\partial}{\partial \bx}\cdot\sum_i\bp_i\Delta_i(\bx)+\left(\bv\cdot\frac{\partial}{\partial \bx}\right)\rho=-\frac{\partial}{\partial \bx}\cdot(\rho \bv)+\left(\bv\cdot\frac{\partial}{\partial \bx}\right)\rho=-\rho\left(\frac{\partial}{\partial \bx}\cdot \bv\right),
\end{aligned}
\end{equation}
and thus have the standard balance of mass:
\begin{equation}
\frac{D\rho}{Dt}=-\rho\left(\frac{\partial}{\partial \bx}\cdot \bv\right).
\end{equation}

\subsection{Balance of linear momentum}
We have
\begin{equation}
\begin{aligned}
\frac{\partial}{\partial t}(\rho \bv)&=\mathcal{F}\sum_i\bp_i\Delta_i(\bx)=\sum_i\left(-\frac{\xi_p}{m}\bp_i+f\bd_i+\sum_j\bbf_{ij}+\alpha_p\bW_i\right)\Delta_i(\bx)+\sum_i\left(\frac{\bp_i}{m}\cdot\frac{\partial}{\partial \bx_i}\right)\bp_i\Delta_i(\bx)\\
&=\rho\left(-\frac{\xi_p}{m}\bv+f\bd+\boldsymbol{\mathcal{W}}\right)+\sum_{ij}\bbf_{ij}\Delta_i(\bx)-\frac{\partial}{\partial \bx}\cdot\sum_i\frac{\bp_i\otimes \bp_i}{m}\Delta_i(\bx).
\end{aligned}
\end{equation}
The second term, which accounts for interparticle forces, may be expressed as a gradient using Noll's formula~\cite{Nol,lehoucq2010translation}
\begin{equation}
\Delta_i(\bx)-\Delta_j(\bx)=-\frac{\partial}{\partial \bx}\cdot(\bx_{ij}b_{ij}).
\end{equation}
where the so-called bond function $b_{ij}$ is defined as
\begin{equation}
b_{ij}=\int_0^1\Delta(\bx-\lambda \bx_i+\bx_{ij})d\lambda,
\end{equation}
with $\bx_{ij}=\bx_i-\bx_j$. Then:
\begin{equation}
\sum_{ij}\bbf_{ij}\Delta_i(\bx)=\frac{1}{2}\sum_{ij}\bbf_{ij}\left(\Delta_i(\bx)-\Delta_j(\bx)\right)=-\frac{1}{2}\sum_{ij}\bbf_{ij}\frac{\partial}{\partial \bx}\cdot(\bx_{ij}b_{ij})=-\frac{\partial}{\partial \bx}\cdot\frac{1}{2}\sum_{ij}b_{ij}\bx_{ij}\otimes \bbf_{ij}.
\end{equation}
The tensor quantity whose gradient is the third term may be expressed as a sum of mean and deviatoric parts:
\begin{equation}
\begin{aligned}
\sum_i\frac{\bp_i\otimes \bp_i}{m}\Delta_i(\bx)&=\sum_i\frac{\left(\bp_i-m\bv\right)\otimes \left(\bp_i-m\bv\right)}{m}\Delta_i(\bx)+\rho \bv\otimes \bv.
\end{aligned}
\end{equation}
Using the balance of mass and then plugging in the above expressions:
\begin{equation}
\begin{aligned}
\rho\frac{D\bv}{Dt}&=\frac{\partial}{\partial t}(\rho \bv)+\frac{\partial}{\partial \bx}\cdot\left(\rho \bv\otimes \bv\right)\\
&=\rho\left(-\frac{\xi_p}{m}\bv+f\bd+\boldsymbol{\mathcal{W}}\right)+\frac{\partial}{\partial \bx}\cdot\left[-\sum_i\frac{\left(\bp_i-m\bv\right)\otimes \left(\bp_i-m\bv\right)}{m}\Delta_i(\bx)-\frac{1}{2}\sum_{ij}b_{ij}\bx_{ij}\otimes \bbf_{ij}\right],
\end{aligned}
\end{equation}
and we have the balance of linear momentum
\begin{equation}
\begin{aligned}
\rho\frac{D\bv}{Dt}&=\rho \mathbf{b}+\frac{\partial}{\partial \bx}\cdot \bT,
\end{aligned}
\end{equation}
with the body force and stress tensor
\begin{equation}
\begin{aligned}
\mathbf{b}&=-\frac{\xi_p}{m}\bv+f\bd+\boldsymbol{\mathcal{W}}\\
\bT&=-\sum_i\frac{\left(\bp_i-m\bv\right)\otimes \left(\bp_i-m\bv\right)}{m}\Delta_i(\bx)-\frac{1}{2}\sum_{ij}b_{ij}\bx_{ij}\otimes \bbf_{ij}.
\end{aligned}
\end{equation}

\subsection{Balance of angular momentum}
The internal angular momentum evolves as follows:
\begin{equation}
\begin{aligned}
\frac{I}{m}\frac{\partial}{\partial t}(\rho w)&=\mathcal{F}\sum_iL_i\Delta_i(\bx)=\sum_i\left(-\frac{\xi_r}{I}L_i+\tau+\sum_j\tau_{ij}+\alpha_rZ_i\right)\Delta_i(\bx)+\sum_i\left(\frac{\bp_i}{m}\cdot\frac{\partial}{\partial \bx_i}\right)L_i\Delta_i(\bx)\\
&=-\frac{\xi_r}{m}\rho w+\rho\frac{\tau}{m}+\rho\mathcal{Z}+\sum_{ij}\tau_{ij}\Delta_i(\bx)-\frac{\partial}{\partial \bx}\cdot\sum_i\frac{\bp_i}{m}L_i\Delta_i(\bx).
\end{aligned}
\end{equation}
As in the case of linear momentum, we apply Noll's formula and split the final term into mean and deviatoric parts:
\begin{equation}
\begin{aligned}
\sum_{ij}\tau_{ij}\Delta_i(x)&=-\frac{\partial}{\partial \bx}\cdot\frac{1}{2}\sum_{ij}b_{ij}\bx_{ij}\tau_{ij}\\
\sum_i\frac{\bp_i}{m}L_i\Delta_i(\bx)&=\sum_i\left(\frac{\bp_i-m\bv}{m}\right)\left(L_i-Iw\right)\Delta_i(\bx)+\frac{I}{m}\rho \bv w.
\end{aligned}
\end{equation}
Using balance of mass and plugging in the above expressions:
\begin{equation}
\begin{aligned}
\frac{I}{m}\rho\frac{Dw}{Dt}&=\frac{I}{m}\frac{\partial}{\partial t}(\rho w)+\frac{I}{m}\frac{\partial}{\partial \bx}\cdot(\bv \rho w)\\
&=\rho\left(-\frac{\xi_r}{I}w+\frac{\tau}{m}+\mathcal{Z}\right)+\frac{\partial}{\partial \bx}\cdot\left[-\frac{1}{2}\sum_{ij}b_{ij}\bx_{ij}\tau_{ij}-\sum_i\left(\frac{\bp_i-m\bv}{m}\right)\left(L_i-Iw\right)\Delta_i(\bx)\right],
\end{aligned}
\end{equation}
and we have the partial balance equation
\begin{equation}
\frac{I}{m}\rho \frac{Dw}{Dt}=\rho G_1+\frac{\partial}{\partial \bx}\cdot \bC_1,
\end{equation}
with body torque and couple stress vector
\begin{equation}
\begin{aligned}
G_1&=-\frac{\xi_r}{m}w+\frac{\tau}{m}+\mathcal{Z}\\
\bC_1&=-\frac{1}{2}\sum_{ij}b_{ij}\bx_{ij}\tau_{ij}-\sum_i\left(\frac{\bp_i-m\bv}{m}\right)\left(L_i-Iw\right).
\end{aligned}
\end{equation}
Next we consider the coarse-graining angular momentum $\rho\boldsymbol{\iota}\cdot\boldsymbol{\Omega}$. First we must evaluate the time-evolution of the moment of inertia $\rho \boldsymbol{\iota}$:
\begin{equation}
\begin{aligned}
\frac{\partial}{\partial t}(\rho\boldsymbol{\iota})&=\mathcal{F}\sum_im\left[(\bx_i-\bx)^2\mathbb{1}-(\bx_i-\bx)\otimes(\bx_i-\bx)\right]\Delta_i(\bx)\\
&=\sum_i\bp_i\cdot\frac{\partial}{\partial \bx_i}\left[(\bx_i-\bx)^2\mathbb{1}-(\bx_i-\bx)\otimes(\bx_i-\bx)\right]\Delta_i(\bx)\\
&\hspace{20pt}+\sum_i\left[(\bx_i-\bx)^2\mathbb{1}-(\bx_i-\bx)\otimes(\bx_i-\bx)\right]\left(\bp_i\cdot\frac{\partial}{\partial \bx_i}\right)\Delta_i(\bx)\\
&=-\frac{\partial}{\partial \bx}\cdot\sum_i\bp_i\otimes\left[(\bx_i-\bx)^2\mathbb{1}-(\bx_i-\bx)\otimes(\bx_i-\bx)\right]\Delta_i(\bx)\\
&=-\frac{\partial}{\partial \bx}\cdot\sum_i\left(\bp_i-m\bv\right)\otimes\left[(\bx_i-\bx)^2\mathbb{1}-(\bx_i-\bx)\otimes(\bx_i-\bx)\right]\Delta_i(\bx)-\frac{\partial}{\partial \bx}\cdot\left(\rho \bv\otimes\boldsymbol{\iota}\right).
\end{aligned}
\end{equation}
Using the mass balance equation:
\begin{equation}
\begin{aligned}
\rho\frac{D\boldsymbol{\iota}}{Dt}&=\frac{\partial}{\partial t}(\rho\boldsymbol{\iota})+\frac{\partial}{\partial \bx}\cdot(\rho \bv\otimes\boldsymbol{\iota}),
\end{aligned}
\end{equation}
so that we have
\begin{equation}
\rho\frac{D\boldsymbol{\iota}}{Dt}=-\frac{\partial}{\partial \bx}\cdot \bY,
\end{equation}
with the coarse-graining moment of inertia flux
\begin{equation}
\bY=\sum_i\left(\bp_i-m\bv\right)\otimes\left[(\bx_i-\bx)^2\mathbb{1}-(\bx_i-\bx)\otimes(\bx_i-\bx)\right]\Delta_i(\bx).
\end{equation}
Now we can examine the coarse-graining angular momentum:
\begin{equation}
\begin{aligned}
\frac{\partial}{\partial t}\left(\rho\boldsymbol{\iota}\cdot\boldsymbol{\Omega}\right)&=\mathcal{F}\sum_i (\bx_i-\bx)\times (\bp_i-m\bv)\Delta_i(\bx)\\
&=\sum_i\frac{\bp_i}{m}\cdot\frac{\partial}{\partial \bx_i}\left[(\bx_i-\bx)\times(\bp_i-m\bv)\Delta_i(\bx)\right]\\
&\hspace{20pt}+\sum_i\left(-\frac{\xi_p}{m}\bp_i+f\bd_i+\sum_j\bbf_{ij}+\alpha_p\bW_i\right)\cdot\frac{\partial}{\partial \bp_i}\left[(\bx_i-\bx)\times(\bp_i-m\bv)\Delta_i(\bx)\right]\\
&=-\frac{\partial}{\partial \bx}\cdot\sum_i\frac{\bp_i}{m}\otimes(\bx_i-\bx)\times(\bp_i-m\bv)\Delta_i(\bx)+\sum_i\left(\bx_i-\bx\right)\times\left(-\frac{\xi_p}{m}\bp_i+f\bd_i+\sum_j\bbf_{ij}+\alpha_p\bW_i\right)\Delta_i(\bx)\\
&=-\frac{\partial}{\partial \bx}\cdot\sum_i\left(\frac{\bp_i}{m}-\bv\right)\otimes(\bx_i-\bx)\times(\bp_i-m\bv)\Delta_i(\bx)-\frac{\partial}{\partial \bx}\cdot\left(\rho \bv\otimes \boldsymbol{\iota}\cdot\boldsymbol{\Omega}\right)\\
&\hspace{20pt}-\frac{\xi_p}{m}\rho\boldsymbol{\iota}\cdot\boldsymbol{\Omega}-\frac{\xi_p}{m}\boldsymbol{\boldsymbol{\delta}}\times\rho \bv+\sum_i\left(\bx_i-\bx\right)\times\left(\alpha_p\bW_i\right)\Delta_i(\bx)\\
&\hspace{20pt}+\sum_{ij}\left(\bx_i-\bx\right)\times f_{ij}\Delta_i(x)+f\sum_i\left(\bx_i-\bx\right)\times \bd_i\Delta_i(\bx).
\end{aligned}
\end{equation}
The interaction term may be expressed as a gradient using Noll's formula:
\begin{equation}
\begin{aligned}
\sum_{ij}\left(\bx_i-\bx\right)\times \bbf_{ij}\Delta_i(\bx)&=\frac{1}{2}\sum_{ij}\left[\left(\bx_i-\bx\right)\times \bbf_{ij}\Delta_i(\bx)-\left(\bx_j-\bx\right)\times \bbf_{ij}\Delta_j(\bx)\right]\\
&=\frac{1}{2}\sum_{ij}\left[\left(\bx_i-\bx\right)\Delta_i(\bx)-\left(\bx_i-\bx\right)\Delta_j(\bx)+\left(\bx_i-\bx\right)\Delta_j(\bx)-\left(\bx_j-\bx\right)\Delta_j(\bx)\right]\times \bbf_{ij}\\
&=\frac{1}{2}\sum_{ij}\left[\left(\bx_i-\bx\right)\left(\Delta_i(\bx)-\Delta_j(\bx)\right)+\left(\bx_i-\bx_j\right)\Delta_j(\bx)\right]\times \bbf_{ij}\\
&=\frac{1}{2}\sum_{ij}\left(\bx_i-\bx\right)\left(\Delta_i(\bx)-\Delta_j(\bx)\right)\times \bbf_{ij}\\
&=-\frac{1}{2}\sum_{ij}\frac{\partial}{\partial \bx}\cdot(\bx_{ij}b_{ij})\left(\bx_i-\bx\right)\times \bbf_{ij}\\
&=-\frac{\partial}{\partial \bx}\cdot\frac{1}{2}\sum_{ij}b_{ij}\bx_{ij}\otimes \left(\bx_i-\bx\right)\times \bbf_{ij},
\end{aligned}
\end{equation}
where we have also used the assumption that the force between particles acts along the ray from one to the other, so that $\bx_i-\bx_j$ is parallel to $\bbf_{ij}$. Using the balances of mass and coarse-graining moment of inertia, the total time derivative of $\boldsymbol{\Omega}$ is now given by
\begin{equation}
\begin{aligned}
\rho\boldsymbol{\iota}\cdot\frac{D\boldsymbol{\Omega}}{Dt}&=\frac{\partial}{\partial t}\left(\rho\boldsymbol{\iota}\cdot\boldsymbol{\Omega}\right)+\left(\bv\cdot\frac{\partial}{\partial \bx}\right)\left(\rho\boldsymbol{\iota}\cdot\boldsymbol{\Omega}\right)-\rho\frac{D\boldsymbol{\iota}}{Dt}\cdot\boldsymbol{\Omega}-\boldsymbol{\iota}\frac{D\rho}{Dt}\cdot\boldsymbol{\Omega}\\
&=\frac{\partial}{\partial t}\left(\rho\boldsymbol{\iota}\cdot\boldsymbol{\Omega}\right)+\left(\bv\cdot\frac{\partial}{\partial \bx}\right)\left(\rho\boldsymbol{\iota}\cdot\boldsymbol{\Omega}\right)+\left(\frac{\partial}{\partial \bx}\cdot \bY\right)\cdot\boldsymbol{\Omega}+\boldsymbol{\iota}\rho\left(\frac{\partial}{\partial \bx}\cdot \bv\right)\cdot\boldsymbol{\Omega}\\
&=\frac{\partial}{\partial t}\left(\rho\boldsymbol{\iota}\cdot\boldsymbol{\Omega}\right)+\left(\frac{\partial}{\partial \bx}\cdot \bY\right)\cdot\boldsymbol{\Omega}+\frac{\partial}{\partial \bx}\cdot\left(\rho \bv\otimes\boldsymbol{\iota}\cdot\boldsymbol{\Omega}\right)\\
&=\frac{\partial}{\partial \bx}\cdot\left[-\sum_i\left(\frac{\bp_i}{m}-\bv\right)\otimes(\bx_i-\bx)\times(\bp_i-m\bv)\Delta_i(\bx)-\frac{1}{2}\sum_{ij}b_{ij}\bx_{ij}\otimes \left(\bx_i-\bx\right)\times \bbf_{ij}\right]\\
&\hspace{20pt}-\frac{\xi_p}{m}\rho\boldsymbol{\iota}\cdot\boldsymbol{\Omega}-\frac{\xi_p}{m}\boldsymbol{\boldsymbol{\delta}}\times\rho \bv+\sum_i\left(\bx_i-\bx\right)\times\left(\alpha_p\bW_i\right)\Delta_i(\bx)+f\sum_i\left(\bx_i-\bx\right)\times \bd_i\Delta_i(\bx)+\left(\frac{\partial}{\partial \bx}\cdot \bY\right)\cdot\boldsymbol{\Omega},
\end{aligned}
\end{equation}
so that we have
\begin{equation}
\rho\boldsymbol{\iota}\cdot\frac{D\boldsymbol{\Omega}}{Dt}=\rho G_2+\frac{\partial}{\partial \bx}\cdot \bC_2+\left(\frac{\partial}{\partial \bx}\cdot \bY\right)\cdot\boldsymbol{\Omega},
\end{equation}
with
\begin{equation}
\begin{aligned}
\rho G_2&=-\frac{\xi_p}{m}\rho\boldsymbol{\iota}\cdot\boldsymbol{\Omega}-\frac{\xi_p}{m}\boldsymbol{\boldsymbol{\delta}}\times\rho \bv+\sum_i\left(x_i-\bx\right)\times\left(\alpha_p\bW_i\right)\Delta_i(\bx)+f\sum_i\left(\bx_i-\bx\right)\times \bd_i\Delta_i(\bx)\\
\bC_2&=-\sum_i\left(\frac{\bp_i}{m}-\bv\right)\otimes(\bx_i-\bx)\times(\bp_i-m\bv)\Delta_i(\bx)-\frac{1}{2}\sum_{ij}b_{ij}\bx_{ij}\otimes \left(\bx_i-\bx\right)\times \bbf_{ij}.
\end{aligned}
\end{equation}
To determine the time-evolution of the moment of linear momentum, we first need to examine the time-evolution of the displacement $\boldsymbol{\boldsymbol{\delta}}$:
\begin{equation}
\begin{aligned}
\frac{\partial}{\partial t}(\rho\boldsymbol{\boldsymbol{\delta}})&=\mathcal{F}\sum_im(\bx_i-\bx)\Delta_i(\bx)=\sum_i\bp_i\cdot\frac{\partial}{\partial \bx_i}\left[(\bx_i-\bx)\Delta_i(\bx)\right]=-\frac{\partial}{\partial \bx}\cdot\sum_i\bp_i\otimes(\bx_i-\bx)\Delta_i(\bx)\\
&=-\frac{\partial}{\partial \bx}\cdot\sum_i(\bp_i-m\bv)\otimes(\bx_i-\bx)\Delta_i(\bx)-\frac{\partial}{\partial \bx}\cdot\sum_im\bv\otimes(\bx_i-\bx)\Delta_i(\bx)\\
&=-\frac{\partial}{\partial \bx}\cdot\sum_i\left(\frac{\bp_i}{m}-\bv\right)\otimes m(\bx_i-\bx)\Delta_i(\bx)-\frac{\partial}{\partial \bx}\cdot\left(\rho \bv\otimes \boldsymbol{\boldsymbol{\delta}}\right).
\end{aligned}
\end{equation}
The total time derivative is then
\begin{equation}
\begin{aligned}
\rho\frac{D\boldsymbol{\boldsymbol{\delta}}}{Dt}&=\frac{\partial}{\partial t}(\rho \boldsymbol{\boldsymbol{\delta}})+\frac{\partial}{\partial \bx}\cdot\left(\rho \bv\otimes\boldsymbol{\boldsymbol{\delta}}\right)=-\frac{\partial}{\partial \bx}\cdot\sum_i\left(\frac{\bp_i}{m}-\bv\right)\otimes m(\bx_i-\bx)\Delta_i(\bx),
\end{aligned}
\end{equation}
and we may write
\begin{equation}
\rho\frac{D\boldsymbol{\boldsymbol{\delta}}}{Dt}=-\frac{\partial}{\partial \bx}\cdot \bR,
\end{equation}
where
\begin{equation}
\bR=\sum_i\left(\frac{\bp_i}{m}-\bv\right)\otimes m(\bx_i-\bx)\Delta_i(\bx).
\end{equation}
Then:
\begin{equation}
\begin{aligned}
\rho\frac{D}{Dt}((\bx+\boldsymbol{\boldsymbol{\delta}})\times \bv)&=\rho\frac{D\boldsymbol{\boldsymbol{\delta}}}{Dt}\times \bv+\rho(\bx+\boldsymbol{\boldsymbol{\delta}})\times\frac{D\bv}{Dt}=-\rho\left(\frac{\partial}{\partial \bx}\cdot \bR\right)\times \bv+(\bx+\boldsymbol{\boldsymbol{\delta}})\times\left(\rho \mathbf{b}+\frac{\partial}{\partial \bx}\cdot \bT\right).
\end{aligned}
\end{equation}
The balance of total angular momentum now follows by combining Eqs. (22), (26), (31), (34), and (37).

\subsection{Balance of energy}
For computational convenience, we will split the energy into parts due to linear momentum, angular momentum, and potential energy, and examine the time-evolution of each of these in turn. For the energy from linear momentum we have:
\begin{equation}
\begin{aligned}
\frac{\partial}{\partial t}(\rho e_p)&=\mathcal{F}\sum_i\frac{\bp_i^2}{2m}\Delta_i(\bx)=\sum_i\frac{\bp_i}{m}\cdot\frac{\partial}{\partial \bx_i}\left[\frac{\bp_i^2}{2m}\Delta_i(\bx)\right]+\sum_i\left(-\frac{\xi_p}{m}\bp_i+f\bd_i+\sum_j\bbf_{ij}+\alpha_p\bW_i\right)\cdot\frac{\partial}{\partial \bp_i}\frac{\bp_i^2}{2m}\Delta_i(\bx)\\
&=-\frac{\partial}{\partial \bx}\cdot\sum_i\frac{\bp_i}{m}\frac{\bp_i^2}{2m}\Delta_i(\bx)+\sum_i\left(-\frac{\xi_p}{m}\bp_i+f\bd_i+\sum_j\bbf_{ij}+\alpha_p\bW_i\right)\cdot\frac{\bp_i}{m}\Delta_i(\bx)\\
&=-\frac{\partial}{\partial \bx}\cdot\sum_i\left(\frac{\bp_i}{m}-\bv\right)\frac{\bp_i^2}{2m}\Delta_i(\bx)-\frac{\partial}{\partial \bx}\cdot\left(\rho \bv e_p\right)\\
&\hspace{20pt}+\sum_i\left(-\frac{\xi_p}{m}\bp_i+f\bd_i+\sum_j\bbf_{ij}+\alpha_p\bW_i\right)\cdot\left(\frac{\bp_i}{m}-\bv\right)\Delta_i(\bx)\\
&\hspace{20pt}+\sum_i\left(-\frac{\xi_p}{m}\bp_i+f\bd_i+\sum_j\bbf_{ij}+\alpha_p\bW_i\right)\Delta_i(\bx)\cdot \bv\\
&=-\frac{\partial}{\partial \bx}\cdot\sum_i\left(\frac{\bp_i}{m}-v\right)\frac{\bp_i^2}{2m}\Delta_i(\bx)-\frac{\partial}{\partial \bx}\cdot\left(\rho \bv e_p\right)+\sum_i\left(-\frac{\xi_p}{m}\bp_i+f\bd_i+\alpha_p\bW_i\right)\cdot\left(\frac{\bp_i}{m}-\bv\right)\Delta_i(\bx)\\
&\hspace{20pt}+\sum_{ij}\bbf_{ij}\cdot\left(\frac{\bp_i}{m}-\bv\right)\Delta_i(\bx)+\sum_{ij}\bbf_{ij}\Delta_i(\bx)\cdot \bv+\rho \mathbf{b}\cdot \bv.
\end{aligned}
\end{equation}
As before, the interaction terms may be re-expressed via Noll's formula:
\begin{equation}
\begin{aligned}
\sum_{ij}\bbf_{ij}\Delta_i(\bx)&=-\frac{\partial}{\partial \bx}\cdot\frac{1}{2}\sum_{ij}b_{ij}\bx_{ij}\otimes \bbf_{ij}=\frac{\partial}{\partial \bx}\cdot \bT^\text{int}\\
\sum_{ij}\bbf_{ij}\cdot\left(\frac{\bp_i}{m}-\bv\right)\Delta_i(\bx)&=\frac{1}{2}\sum_{ij}\left[\bbf_{ij}\cdot\left(\frac{\bp_i}{m}-\bv\right)\Delta_i(\bx)-\bbf_{ij}\cdot\left(\frac{\bp_j}{m}-\bv\right)\Delta_j(\bx)\right]\\
&=\frac{1}{2}\sum_{ij}\bbf_{ij}\cdot\left[\left(\frac{\bp_i}{m}-\bv\right)\Delta_i(\bx)+\left(\frac{\bp_i}{m}-\bv\right)\Delta_j(\bx)-\left(\frac{\bp_i}{m}-\bv\right)\Delta_j(\bx)-\left(\frac{\bp_j}{m}-\bv\right)\Delta_j(\bx)\right]\\
&=\frac{1}{2}\sum_{ij}\bbf_{ij}\cdot\left[\left(\frac{\bp_i}{m}-\bv\right)\left(\Delta_i(\bx)-\Delta_j(\bx)\right)+\left(\frac{\bp_i}{m}-\bv\right)\Delta_j(\bx)-\left(\frac{\bp_j}{m}-\bv\right)\Delta_j(\bx)\right]\\
&=\frac{1}{2}\sum_{ij}\bbf_{ij}\cdot\left(\frac{\bp_i}{m}-\bv\right)\left(\Delta_i(\bx)-\Delta_j(\bx)\right)+\frac{1}{2}\sum_{ij}\bbf_{ij}\cdot\left[\left(\frac{\bp_i}{m}-\bv\right)\Delta_j(\bx)-\left(\frac{\bp_j}{m}-\bv\right)\Delta_j(\bx)\right]\\
&=-\frac{1}{2}\sum_{ij}\bbf_{ij}\cdot\left(\frac{\bp_i}{m}-\bv\right)\frac{\partial}{\partial \bx}\cdot(\bx_{ij}b_{ij})+\frac{1}{2}\sum_{ij}\bbf_{ij}\cdot\left(\frac{\bp_i}{m}-\frac{\bp_j}{m}\right)\Delta_j(\bx)\\
&=-\frac{\partial}{\partial \bx}\cdot\frac{1}{2}\sum_{ij}b_{ij}\left(\bx_{ij}\otimes \bbf_{ij}\right)\cdot\left(\frac{\bp_i}{m}-\bv\right)-\frac{1}{2}\sum_{ij}\left[(\bx_{ij}b_{ij})\cdot\left(\frac{\partial}{\partial \bx}\bv\right)\cdot \bbf_{ij}\right]\\
&\hspace{20pt}+\frac{1}{2}\sum_{ij}\bbf_{ij}\cdot\left(\frac{\bp_i}{m}-\frac{\bp_j}{m}\right)\Delta_j(\bx)\\
&=-\frac{\partial}{\partial \bx}\cdot\frac{1}{2}\sum_{ij}b_{ij}\left(\bx_{ij}\otimes \bbf_{ij}\right)\cdot\left(\frac{\bp_i}{m}-\bv\right)-\left(\frac{\partial}{\partial \bx}\bv\right)^T:\frac{1}{2}\sum_{ij}\bx_{ij}b_{ij}\otimes \bbf_{ij}\\
&\hspace{20pt}+\frac{1}{2}\sum_{ij}\bbf_{ij}\cdot\left(\frac{\bp_i}{m}-\frac{\bp_j}{m}\right)\Delta_j(\bx)\\
&=-\frac{\partial}{\partial \bx}\cdot\frac{1}{2}\sum_{ij}b_{ij}\left(\bx_{ij}\otimes \bbf_{ij}\right)\cdot\left(\frac{\bp_i}{m}-\bv\right)+\left(\frac{\partial}{\partial \bx}\bv\right)^T:\bT^\text{int}+\frac{1}{2}\sum_{ij}\bbf_{ij}\cdot\left(\frac{\bp_i}{m}-\frac{\bp_j}{m}\right)\Delta_j(\bx).
\end{aligned}
\end{equation}
We can also expand:
\begin{equation}
\begin{aligned}
\sum_i\left(\frac{\bp_i}{m}-\bv\right)\frac{\bp_i^2}{2m}\Delta_i(\bx)&=\sum_i\left(\frac{\bp_i}{m}-\bv\right)\frac{\left(\bp_i-m\bv\right)^2}{2m}\Delta_i(\bx)+\sum_i\frac{\left(\bp_i-m\bv\right)\otimes\left(\bp_i-m\bv\right)}{m}\Delta_i(\bx)\cdot \bv\\
&=\sum_i\left(\frac{\bp_i}{m}-\bv\right)\frac{\left(\bp_i-m\bv\right)^2}{2m}\Delta_i(\bx)-\bT^\text{free}\cdot \bv.
\end{aligned}
\end{equation}
Plugging these in:
\begin{equation}
\begin{aligned}
\rho\frac{De_p}{Dt}&=\frac{\partial}{\partial t}(\rho e_p)+\frac{\partial}{\partial \bx}\cdot(\rho \bv e_p)\\
&=\rho \mathbf{b}\cdot \bv+\frac{\partial}{\partial \bx}\cdot(\bT\cdot \bv)+\sum_i\left(-\frac{\xi_p}{m}\bp_i+f\be_i+\alpha_p\bW_i\right)\cdot\left(\frac{\bp_i}{m}-\bv\right)\Delta_i(\bx)\\
&\hspace{20pt}-\frac{\partial}{\partial \bx}\cdot\left[\sum_i\left(\frac{\bp_i}{m}-\bv\right)\frac{\left(\bp_i-m\bv\right)^2}{2m}\Delta_i(\bx)+\frac{1}{2}\sum_{ij}b_{ij}\left(\bx_{ij}\otimes \bbf_{ij}\right)\cdot\left(\frac{\bp_i}{m}-\bv\right)\right]+\frac{1}{2}\sum_{ij}\bbf_{ij}\cdot\left(\frac{\bp_i}{m}-\frac{\bp_j}{m}\right)\Delta_j(\bx).
\end{aligned}
\end{equation}
The third term is a body heating term, the fourth a heat flux, and the last a term accounting for exchange between kinetic and potential energy.\\
\\
For the energy due to angular momentum we have:
\begin{equation}
\begin{aligned}
\frac{\partial}{\partial t}(\rho e_r)&=\mathcal{F}\sum_i\frac{L_i^2}{2I}\Delta_i(x)=\sum_i\frac{\bp_i}{m}\cdot\frac{\partial}{\partial \bx_i}\left[\frac{L_i^2}{2I}\Delta_i(\bx)\right]+\sum_i\left(-\frac{\xi_r}{I}L_i+\tau+\sum_j\tau_{ij}+\alpha_rZ_i\right)\frac{\partial}{\partial L_i}\frac{L_i^2}{2I}\Delta_i(\bx)\\
&=-\frac{\partial}{\partial \bx}\cdot\sum_i\frac{\bp_i}{m}\frac{L_i^2}{2I}\Delta_i(\bx)+\sum_i\left(-\frac{\xi_r}{I}L_i+\tau+\sum_j\tau_{ij}+\alpha_rZ_i\right)\frac{L_i}{I}\Delta_i(\bx)\\
&=-\frac{\partial}{\partial \bx}\cdot\sum_i\frac{\bp_i}{m}\frac{L_i^2}{2I}\Delta_i(\bx)+\sum_i\left(-\frac{\xi_r}{I}L_i+\tau+\sum_j\tau_{ij}+\alpha_rZ_i\right)\left(\frac{L_i}{I}-w\right)\Delta_i(\bx)\\
&\hspace{20pt}+\sum_i\left(-\frac{\xi_r}{I}L_i+\tau+\sum_j\tau_{ij}+\alpha_rZ_i\right)\Delta_i(\bx)w\\
&=-\frac{\partial}{\partial \bx}\cdot\sum_i\frac{\bp_i}{m}\frac{L_i^2}{2I}\Delta_i(\bx)+\sum_i\left(-\frac{\xi_r}{I}L_i+\tau+\sum_j\tau_{ij}+\alpha_rZ_i\right)\left(\frac{L_i}{I}-w\right)\Delta_i(\bx)\\
&\hspace{20pt}+\frac{\tau}{m}\rho w+\rho\mathcal{Z}w-\frac{\xi_r}{I}\frac{I}{m}\rho w^2+\sum_{ij}\tau_{ij}\Delta_i(\bx)w.
\end{aligned}
\end{equation}
The first term can be rewritten as follows:
\begin{equation}
\begin{aligned}
\sum_i\frac{\bp_i}{m}\frac{L_i^2}{2I}\Delta_i(\bx)&=\sum_i\left(\frac{\bp_i}{m}-\bv\right)\frac{\left(L_i-Iw\right)\left(L_i-Iw\right)}{2I}\Delta_i(\bx)+\sum_i\left(\frac{\bp_i}{m}-\bv\right)\left(L_i-Iw\right)\Delta_i(\bx)w+\rho \bv\sum_i\frac{L_i^2}{2I}\Delta_i(\bx)\\
&=\sum_i\left(\frac{\bp_i}{m}-\bv\right)\frac{\left(L_i-Iw\right)\left(L_i-Iw\right)}{2I}\Delta_i(\bx)-\bC_1^\text{free}w+\rho \bv e_r,
\end{aligned}
\end{equation}
and the interaction terms as follows, identically to the linear momentum energy case above:
\begin{equation}
\begin{aligned}
\sum_{ij}\tau_{ij}\Delta_i(\bx)w&=-\left(\frac{\partial}{\partial \bx}\cdot\frac{1}{2}\sum_{ij}b_{ij}\bx_{ij}\tau_{ij}\right)w=\left(\frac{\partial}{\partial \bx}\cdot \bC_1^\text{int}\right)w\\
\sum_{ij}\tau_{ij}\left(\frac{L_i}{I}-w\right)\Delta_i(\bx)&=-\frac{\partial}{\partial \bx}\cdot\frac{1}{2}\sum_{ij}b_{ij}\bx_{ij} \tau_{ij}\left(\frac{L_i}{I}-w\right)+\left(\frac{\partial}{\partial \bx}w\right)\cdot \bC_1^\text{int}+\frac{1}{2}\sum_{ij}\tau_{ij}\left(\frac{L_i}{I}-\frac{L_j}{I}\right)\Delta_j(\bx).
\end{aligned}
\end{equation}
Plugging these in:
\begin{equation}
\begin{aligned}
\rho\frac{De_r}{Dt}&=\frac{\partial}{\partial t}(\rho e_r)+\frac{\partial}{\partial \bx}\cdot(\rho \bv e_r)\\
&=\rho G_1w+\frac{\partial}{\partial \bx}\cdot\left(\bC_1w\right)+\sum_i\left(-\frac{\xi_r}{I}L_i+\tau+\alpha_rZ_i\right)\left(\frac{L_i}{I}-w\right)\Delta_i(\bx)\\
&\hspace{20pt}-\frac{\partial}{\partial \bx}\cdot\left[\sum_i\left(\frac{\bp_i}{m}-\bv\right)\frac{\left(L_i-Iw\right)^2}{2I}\Delta_i(\bx)+\frac{1}{2}\sum_{ij}b_{ij}\bx_{ij} \tau_{ij}\left(\frac{L_i}{I}-w\right)\right]+\frac{1}{2}\sum_{ij}\tau_{ij}\left(\frac{L_i}{I}-\frac{L_j}{I}\right)\Delta_j(\bx).
\end{aligned}
\end{equation}
The interpretation of these terms is the same as for the energy due to linear momentum.\\
\\
For potential energy, we have:
\begin{equation}
\begin{aligned}
\frac{\partial}{\partial t}(\rho e_u)&=\mathcal{F}\sum_{ij}\frac{u_{ij}}{2}\Delta_i(\bx)=\sum_{ijk}\frac{\bp_k}{m}\cdot\frac{\partial}{\partial \bx_k}\left[\frac{u_{ij}}{2}\Delta_i(\bx)\right]+\sum_{ijk}\frac{L_k}{I}\frac{\partial}{\partial \theta_k}\left[\frac{u_{ij}}{2}\Delta_i(\bx)\right]\\
&=\sum_{ijk}\frac{u_{ij}}{2}\frac{\bp_k}{m}\cdot\frac{\partial}{\partial \bx_k}\Delta_i(\bx)+\frac{1}{2}\sum_{ijk}\frac{\bp_k}{m}\cdot\frac{\partial u_{ij}}{\partial \bx_k}\Delta_i(\bx)+\frac{1}{2}\sum_{ijk}\frac{L_k}{I}\frac{\partial u_{ij}}{\partial \theta_k}\Delta_i(\bx)\\
&=-\frac{\partial}{\partial \bx}\cdot\sum_{ij}\frac{\bp_i}{m}\frac{u_{ij}}{2}\Delta_i(\bx)+\frac{1}{2}\sum_{ijk}\frac{\bp_k}{m}\cdot\left(\boldsymbol{\boldsymbol{\delta}}_{ik}\frac{\partial u_{ij}}{\partial \bx_k}+\boldsymbol{\boldsymbol{\delta}}_{jk}\frac{\partial u_{ij}}{\partial \bx_k}\right)\Delta_i(\bx)+\frac{1}{2}\sum_{ijk}\frac{L_k}{I}\left(\boldsymbol{\boldsymbol{\delta}}_{ik}\frac{\partial u_{ij}}{\partial \theta_k}+\boldsymbol{\boldsymbol{\delta}}_{jk}\frac{\partial u_{ij}}{\partial \theta_k}\right)\Delta_i(\bx)\\
&=-\frac{\partial}{\partial \bx}\cdot\sum_{ij}\left(\frac{\bp_i}{m}-\bv\right)\frac{u_{ij}}{2}\Delta_i(\bx)-\frac{\partial}{\partial \bx}\cdot\left(\rho \bv e_u\right)-\frac{1}{2}\sum_{ij}\left(\frac{\bp_i}{m}-\frac{\bp_j}{m}\right)\cdot \bbf_{ij}\Delta_i(\bx)-\frac{1}{2}\sum_{ij}\left(\frac{L_i}{I}-\frac{L_j}{I}\right) \tau_{ij}\Delta_i(\bx).
\end{aligned}
\end{equation}
Then:
\begin{equation}
\begin{aligned}
\rho\frac{De_u}{Dt}&=\frac{\partial}{\partial t}(\rho e_u)+\frac{\partial}{\partial \bx}\cdot(\rho \bv e_u)\\
&=-\frac{\partial}{\partial \bx}\cdot\sum_{ij}\left(\frac{\bp_i}{m}-\bv\right)\frac{u_{ij}}{2}\Delta_i(\bx)-\frac{1}{2}\sum_{ij}\left(\frac{\bp_i}{m}-\frac{\bp_j}{m}\right)\cdot \bbf_{ij}\Delta_i(\bx)-\frac{1}{2}\sum_{ij}\left(\frac{L_i}{I}-\frac{L_j}{I}\right) \tau_{ij}\Delta_i(\bx).
\end{aligned}
\end{equation}
Now we may write the total energy balance:
\begin{equation}
\begin{aligned}
\rho\frac{De}{Dt}&=\rho \mathbf{b}\cdot \bv+\frac{\partial}{\partial \bx}\cdot(\bT\cdot \bv)+\rho G_1w+\frac{\partial}{\partial \bx}\cdot\left(\bC_1w\right)+\rho r-\frac{\partial}{\partial \bx}\cdot \bq,
\end{aligned}
\end{equation}
where the body heating $\rho r$ and heat flux vector $\bq$ are given by
\begin{equation}
\begin{aligned}
\rho r&=\sum_i\left(-\frac{\xi_p}{m}\bp_i+f\bd_i+\alpha_p\bW_i\right)\cdot\left(\frac{\bp_i}{m}-\bv\right)\Delta_i(\bx)+\sum_i\left(-\frac{\xi_r}{I}L_i+\tau+\alpha_rZ_i\right)\left(\frac{L_i}{I}-w\right)\Delta_i(\bx)\\
\bq&=\sum_i\left(\frac{\bp_i}{m}-\bv\right)\left[\frac{\left(\bp_i-m\bv\right)^2}{2m}+\frac{\left(L_i-Iw\right)^2}{2I}+\sum_j\frac{u_{ij}}{2}\right]\Delta_i(\bx)+\frac{1}{2}\sum_{ij}b_{ij}\left[\left(\bx_{ij}\otimes \bbf_{ij}\right)\cdot\left(\frac{\bp_i}{m}-\bv\right)+\bx_{ij} \tau_{ij}\left(\frac{L_i}{I}-w\right)\right].
\end{aligned}
\end{equation}

\section{Director Field Dynamics}
\subsection{Underdamped Case}
\noindent For the linearly and angularly underdamped dynamics used to derive the balance laws, we have:
\begin{equation}
\begin{aligned}
\frac{\partial}{\partial t}(\rho \bd)&=\mathcal{F}\sum_i\bd_i\Delta_i(\bx)=\sum_i\frac{\bp_i}{m}\cdot\frac{\partial}{\partial \bx_i}\left[\bd_i\Delta_i(\bx)\right]+\sum_i\frac{L_i}{I}\frac{\partial}{\partial \theta_i}\bd_i\Delta_i(\bx)\\
&=-\frac{\partial}{\partial \bx}\cdot\sum_i\frac{\bp_i}{m}\otimes \bd_i\Delta_i(\bx)+\sum_i\frac{L_i}{I}{\bd}_i^\perp\Delta_i(\bx)\\
&=-\frac{\partial}{\partial \bx}\cdot\sum_i\left(\frac{\bp_i}{m}-\bv\right)\otimes \bd_i\Delta_i(\bx)-\frac{\partial}{\partial \bx}\cdot\sum_i\bv\otimes \bd_i\Delta_i(\bx)+\sum_i\left(\frac{L_i}{I}-w\right){\bd}^\perp_i\Delta_i(\bx)+\sum_iw {\bd}^\perp_i\Delta_i(\bx)\\
&=-\frac{\partial}{\partial \bx}\cdot\sum_i\left(\frac{\bp_i}{m}-\bv\right)\otimes \bd_i\Delta_i(\bx)-\frac{\partial}{\partial \bx}\cdot\left(\rho \bv\otimes \bd\right)+\sum_i\left(\frac{L_i}{I}-w\right){\bd}^\perp_i\Delta_i(\bx)+\rho w {\bd}^\perp,
\end{aligned}
\end{equation}
where the $\perp$ indicates rotation by $\pi/2$ counterclockwise. Then we can evaluate the total time derivative:
\begin{equation}
\begin{aligned}
\rho\frac{D\bd}{Dt}&=\frac{\partial}{\partial t}(\rho \bd)+\frac{\partial}{\partial \bx}\cdot(\rho \bv\otimes \bd)\\
&=-\frac{\partial}{\partial \bx}\cdot\sum_i\left(\frac{\bp_i}{m}-\bv\right)\otimes \bd_i\Delta_i(\bx)+\sum_i\left(\frac{L_i}{I}-w\right){\bd}^\perp_i\Delta_i(\bx)+\rho w {\bd}^\perp\\
&=-\frac{\partial}{\partial \bx}\cdot\sum_i\left(\frac{\bp_i}{m}-\bv\right)\otimes \bd_i\Delta_i(\bx)+\sum_i\left(\frac{L_i}{I}-w\right)\left({\bd}^\perp_i-{\bd}^\perp\right)\Delta_i(\bx)+\rho w {\bd}^\perp.
\end{aligned}
\end{equation}

\subsection{Overdamped and Averaged Case}
\noindent In this case, we have the angular dynamics
\begin{equation}
\frac{d\theta_i}{dt}=\mu\tau+\mu\sum_j\tau_{ij}+\mu\alpha_rZ_i(t),
\end{equation}
where $\mu$ is an angular mobility. We also define fields in terms of noise averages:
\begin{equation}
\begin{aligned}
\bar{\rho}&=\brak{\sum_im\Delta_i(\bx)}\hspace{20pt}\bar{\rho} \bar{\bv}=\brak{\sum_im\frac{d\bx_i}{dt}\Delta_i(\bx)}\hspace{20pt}\bar{\rho} \bar{\bd}=\brak{\sum_i\bd_i\Delta_i(\bx)},
\end{aligned}
\end{equation}
where we use $\displaystyle{\frac{d\bx_i}{dt}}$ in place of $\bp_i$ in order to allow for either over- or under-damped linear dynamics. The director density dynamics is then
\begin{equation}
\begin{aligned}
\frac{\partial}{\partial t}(\bar{\rho} \bar{\bd})&=\brak{\sum_i\frac{d\bx_i}{dt}\cdot\frac{\partial}{\partial \bx_i}\left[\bd_i\Delta_i(\bx)\right]}+\brak{\sum_i\frac{d\theta_i}{dt}\frac{\partial}{\partial \theta_i}\left[\bd_i\Delta_i(\bx)\right]}\\
&=-\frac{\partial}{\partial \bx}\cdot\brak{\sum_i\frac{d\bx_i}{dt}\otimes \bd_i\Delta_i(\bx)}+\brak{\sum_i\left(\mu\tau+\mu\sum_j\tau_{ij}+\mu\alpha_rZ_i\right)\bd^\perp_i\Delta_i(\bx)}\\
&=-\frac{\partial}{\partial \bx}\cdot\brak{\sum_i\frac{d\bx_i}{dt}\otimes \bd_i\Delta_i(\bx)}+\mu\tau\bar{\rho} \bar{\bd}^\perp+\mu\brak{\sum_{ij}\tau_{ij}\bd^\perp_i\Delta_i(\bx)}+\mu\alpha_r\brak{\sum_iZ_i\bd^\perp_i\Delta_i(\bx)}.
\end{aligned}
\end{equation}
We need the mass balance in this form as well:
\begin{equation}
\begin{aligned}
\frac{D\bar{\rho}}{Dt}&=\frac{\partial\bar{\rho}}{\partial t}+\bar{\bv}\cdot\frac{\partial\bar{\rho}}{\partial \bx}=\brak{\sum_im\frac{d\bx_i}{dt}\cdot\frac{\partial}{\partial \bx_i}\Delta_i(\bx)}+\bar{\bv}\cdot\frac{\partial\bar{\rho}}{\partial \bx}=-\frac{\partial}{\partial \bx}\cdot \left(\bar{\rho} \bar{\bv}\right)+\bar{\bv}\cdot\frac{\partial\bar{\rho}}{\partial \bx}=-\bar{\rho}\left(\frac{\partial}{\partial \bx}\cdot \bar{\bv}\right),
\end{aligned}
\end{equation}
as in the non-averaged case. Now the total time derivative of the director field is
\begin{equation}
\begin{aligned}
\bar{\rho}\frac{D\bar{\bd}}{Dt}&=-\frac{\partial}{\partial \bx}\cdot\brak{\sum_i\left(\frac{d\bx_i}{dt}-\bar{\bv}\right)\otimes \bd_i\Delta_i(\bx)}+\mu\tau\bar{\rho} \bar{\bd}^\perp+\mu\brak{\sum_{ij}\tau_{ij}\bar{\bd}^\perp_i\Delta_i(\bx)}+\mu\alpha_r\brak{\sum_iZ_i\bd^\perp_i\Delta_i(\bx)}.
\end{aligned}
\end{equation}
Picking some initial time $t=0$ far in the past, we can write
\begin{equation}
\begin{aligned}
\bd_i(t)&=\bd_i(0)+\int_0^t\frac{d\bd_i}{dt'}dt'=\bd_i(0)+\int_0^t\frac{d\bd_i}{d\theta_i}\frac{d\theta_i}{dt'}dt'=\bd_i(0)+\mu\tau\int_0^t\bd^\perp_idt'+\mu\sum_j\int_0^t\bd^\perp_i\tau_{ij}dt'+\mu\alpha_r\int_0^t\bd^\perp_iZ_i(t')dt'.
\end{aligned}
\end{equation}
Then:
\begin{equation}
\begin{aligned}
\brak{\mu\alpha_r\sum_iZ_i\bd^\perp_i\Delta_i(\bx)}&=\brak{\mu\alpha_r\sum_iZ_i(t)\left[-\mu\alpha_r\int_0^t\bd_iZ_i(t')dt'\right]\Delta_i(\bx)}\\
&=-\frac{1}{2}\mu^2\alpha_r^2\bar{\rho}\bar{\bd},
\end{aligned}
\end{equation}
\textcolor{black}{where use is made of the Stratonovich convention.} Now we finally have
\begin{equation}
\begin{aligned}
\bar{\rho}\frac{D\bar{\bd}}{Dt}&=-\frac{1}{2}\mu^2\alpha_r^2\rho \bar{\bd}+\mu\tau\bar{\rho} \bar{\bd}^\perp+\boldsymbol{\mathcal{A}}-\frac{\partial}{\partial \bx}\cdot \bJ_d,
\end{aligned}
\end{equation}
where
\begin{equation}
\bJ_d=\brak{\sum_i\left(\frac{d\bx_i}{dt}-\bv\right)\otimes \bd_i\Delta_i(\bx)}
\end{equation}
\begin{equation}
\boldsymbol{\mathcal{A}}=\mu\brak{\sum_{ij}\tau_{ij}\bd^\perp_i\Delta_i(\bx)}.
\end{equation}

\section{Virial Theorems}\label{app:Virial}
In this appendix, we provide derivations of the atomistic and continuum virial theorems. Both are well-known, and in particular the atomistic version may be found in many introductions to classical mechanics, e.g. \cite{goldstein2002classical}.
\subsection{Atomistic}
For a system of particles with positions $\mathbf{x}_i$, the second moment of the mass distribution takes the form
\begin{equation}
\mathbf{R}_2=\sum_i(\mathbf{x}_i-\mathbf{x}_\text{cm})\otimes (\mathbf{x}_i-\mathbf{x}_\text{cm}),
\end{equation}
with $\mathbf{x}_\text{cm}$ the center of mass. Taking the second time-derivative, we find
\begin{equation}
\begin{aligned}
\frac{d^2}{dt^2}\mathbf{R}_2&=2\sum_i\left[\left(\frac{d^2\mathbf{x}_i}{dt^2}-\frac{d^2\mathbf{x}_\text{cm}}{dt^2}\right)\otimes (\mathbf{x}_i-\mathbf{x}_\text{cm})+\left(\frac{d\mathbf{x}_i}{dt}-\frac{d\mathbf{x}_\text{cm}}{dt}\right)\otimes \left(\frac{d\mathbf{x}_i}{dt}-\frac{d\mathbf{x}_\text{cm}}{dt}\right)\right]^\text{sym},
\end{aligned}
\end{equation}
where the superscript denotes taking the symmetric part of the tensor. Assuming that the system is in a steady state, then we should have $d^2\mathbf{R}_2/dt^2=0$, where brackets indicate time- or ensemble-averaging. Moreover, the center of mass should be approximately constant, and we may as well choose coordinates such that $\mathbf{x}_\text{cm}=0$. Then
\begin{equation}
\begin{aligned}
0&=\brak{\sum_i\left[\frac{d^2\mathbf{x}_i}{dt^2}\otimes\mathbf{x}_i+\frac{d\mathbf{x}_i}{dt}\otimes\frac{d\mathbf{x}_i}{dt}\right]^\text{sym}}.\label{tensorvirial}
\end{aligned}
\end{equation}
Using the equations of motion $d\mathbf{x}_i/dt=\mathbf{p}_i/m$ and $d\mathbf{p}_i/dt=\mathbf{f}_i$ and taking the trace of Eq. (\ref{tensorvirial}), we find
\begin{equation}
\begin{aligned}
\brak{\sum_i\mathbf{f}_i\cdot\mathbf{x}_i+\sum_i\frac{\mathbf{p}_i\cdot \mathbf{p}_i}{m}}&=0.
\end{aligned}
\end{equation}

\subsection{Continuum}
Proceeding in the same manner as in the atomistic case, we define the second moment of the mass distribution:
\begin{equation}
\begin{aligned}
\mathbf{R}_2&=\int_{\mathbb{R}^d}\rho\left(\mathbf{x}-\mathbf{x}_\text{cm}\right)\otimes\left(\mathbf{x}-\mathbf{x}_\text{cm}\right)dv,
\end{aligned}
\end{equation}
where $\mathbf{x}_\text{cm}$ is again the center of mass. We consider the fields to be defined on all of space $\mathbb{R}^d$, and to vanish at infinity. In steady-state, we again set $\mathbf{x}_\text{cm}=0$. Then, taking the second time-derivative and using the continuum equations of motion
\begin{equation}
\begin{aligned}
\frac{\partial\rho}{\partial t}&=-\frac{\partial}{\partial\mathbf{x}}\cdot(\rho\mathbf{v})\\
\rho\frac{\partial\mathbf{v}}{\partial t}&=\rho\mathbf{b}+\frac{\partial}{\partial\mathbf{x}}\cdot\mathbf{T}-\rho\mathbf{v}\cdot\frac{\partial}{\partial\mathbf{x}}\mathbf{v}
\end{aligned}
\end{equation}
we find
\begin{equation}
\begin{aligned}
\frac{d^2}{dt^2}\mathbf{R}_2&=\frac{d^2}{dt^2}\int_{\mathbb{R}^d}\rho\mathbf{x}\otimes\mathbf{x}\,dv\\
&=\frac{d}{dt}\int_{\mathbb{R}^d}\frac{\partial\rho}{\partial t}\mathbf{x}\otimes\mathbf{x}\,dv\\
&=-\frac{d}{dt}\int_{\mathbb{R}^d}\left[\frac{\partial}{\partial\mathbf{x}}\cdot(\rho\mathbf{v})\right]\mathbf{x}\otimes\mathbf{x}\,dv\\
&=-\int_{\mathbb{R}^d}\left[\frac{\partial}{\partial\mathbf{x}}\cdot\left(\frac{\partial\rho}{\partial t}\mathbf{v}+\rho\frac{\partial\mathbf{v}}{\partial t}\right)\right]\mathbf{x}\otimes\mathbf{x}\,dv\\
&=\int_{\mathbb{R}^d}\left(\frac{\partial\rho}{\partial t}\mathbf{v}+\rho\frac{\partial\mathbf{v}}{\partial t}\right)\cdot\frac{\partial}{\partial\mathbf{x}}\left(\mathbf{x}\otimes\mathbf{x}\right)dv\\
&=2\int_{\mathbb{R}^d}\left(\frac{\partial\rho}{\partial t}\mathbf{v}+\rho\frac{\partial\mathbf{v}}{\partial t}\right)\otimes\mathbf{x}\,dv\\
&=2\int_{\mathbb{R}^d}\left(-\left(\frac{\partial}{\partial\mathbf{x}}\cdot(\rho\mathbf{v})\right)\mathbf{v}+\rho\mathbf{b}+\frac{\partial}{\partial\mathbf{x}}\cdot\mathbf{T}-\rho\mathbf{v}\cdot\frac{\partial}{\partial\mathbf{x}}\mathbf{v}\right)\otimes\mathbf{x}\,dv\\
&=2\int_{\mathbb{R}^d}\left(-\frac{\partial}{\partial\mathbf{x}}\cdot(\rho\mathbf{v}\otimes\mathbf{v})+\rho\mathbf{b}+\frac{\partial}{\partial\mathbf{x}}\cdot\mathbf{T}\right)\otimes\mathbf{x}\,dv\\
&=-2\int_{\mathbb{R}^d}\frac{\partial}{\partial\mathbf{x}}\cdot(\rho\mathbf{v}\otimes\mathbf{v})\otimes\mathbf{x}\,dv+d\int_{\mathbb{R}^d}\rho\mathbf{b}\otimes\mathbf{x}\,dv+2\int_{\mathbb{R}^d}\left(\frac{\partial}{\partial\mathbf{x}}\cdot\mathbf{T}\right)\otimes\mathbf{x}\,dv\\
&=2\int_{\mathbb{R}^d}\rho\mathbf{v}\otimes\mathbf{v}\,dv+2\int_{\mathbb{R}^d}\rho\mathbf{b}\otimes\mathbf{x}\,dv-2\int_{\mathbb{R}^d}\mathbf{T}\,dv,
\end{aligned}
\end{equation}
where we have repeatedly integrated by parts, assuming that all fields vanish at infinity. Setting $d^2\mathbf{R}_2/dt^2=0$ and taking the trace, we find the continuum version of the virial theorem to be
\begin{equation}
\int_{\mathbb{R}^d}\rho\mathbf{v} \cdot \mathbf{v} dv+\int_{\mathbb{R}^d}\rho\mathbf{b}\cdot\mathbf{x}\,dv-\int_{\mathbb{R}^d}\tr(\mathbf{T})dv=0.
\end{equation}

    \setlength{\bibsep}{10pt}
    \linespread{1}\selectfont

\end{document}